\documentclass[reprint,nofootinbib]{revtex4}
\usepackage{graphicx}  
\usepackage{dcolumn}   
\usepackage{bm}        
\usepackage{amssymb}
\usepackage{multirow}
\usepackage{amsmath}
\usepackage{cases}
\usepackage{placeins}
\usepackage{textcomp}

\usepackage{blindtext}
\usepackage{url}
\usepackage{graphicx}
\usepackage{url}
\usepackage{tikz}
\usepackage[bf]{caption}
\usepackage{amsmath}
\usepackage{cases} 
\usepackage{multirow}

\usepackage{epsfig}
\usepackage{makecell}
\usepackage{amsmath}
\usepackage{amsfonts}
\usepackage{amssymb}
\usepackage{mathtools}
\usepackage{url}
\usepackage{subfigure}
\usepackage{hhline}
\usepackage{color}
\usepackage{array,multirow}
\usepackage{bm}
\usepackage{makecell}
\usepackage{epsfig}
\usepackage{amsmath}
\usepackage{amsfonts}
\usepackage{amssymb}
\usepackage{url}
\usepackage{hyperref}
\usepackage{subfigure}
\usepackage{hhline}
\usepackage{color}
\usepackage{bm}
\usepackage{cancel}
\usepackage{mathtools}

\newcommand{\beqa}{\begin{eqnarray}}
\newcommand{\eeqa}{\end{eqnarray}}
\newcommand{\beq}{\begin{equation}}
\newcommand{\eeq}{\end{equation}}

\newcommand{\nn}{\nonumber}
\newcommand{\bmt}{\begin{pmatrix}}
\newcommand{\emt}{\end{pmatrix}}
\usepackage[toc,page]{appendix}
\usepackage{comment}
\newcommand{\be}{\begin{equation}}
\newcommand{\ee}{\end{equation}}
\newcommand{\bea}{\begin{eqnarray}}
\newcommand{\eea}{\end{eqnarray}}

\def\Lb{{\Lambda_b}}
\def\Lst{{\Lambda^\ast}}
\def\mLb{{m_{\Lambda_b}}}

\def\mL{{m_{\Lambda^\ast}}}

\def\cl{{\cos\theta_\ell}}
\def\ccl{{\cos^2\theta_\ell}}
\def\sl{{\sin\theta_\ell}}
\def\ssl{{\sin^2\theta_\ell}}
\def\cLst{{\cos\theta_{\Lst}}}
\def\ccLst{{\cos^2\theta_{\Lst}}}
\def\sLst{{\sin\theta_{\Lst}}}
\def\ssLst{{\sin^2\theta_{\Lst}}}

\definecolor{schrift}{RGB}{120,0,0}

\begin{document}

\title{New physics analysis of $\Lambda_b\to (\Lambda^*(\to pK^-), \Lambda(\to p\pi))({\mu}^{+}\mu^{-},\,\nu\bar{\nu})$  baryonic decays under SMEFT framework}
\author{Nilakshi~Das${}$}
\email{nilakshi\_rs@phy.nits.ac.in} 
\author{Rupak~Dutta${}$}
\email{rupak@phy.nits.ac.in}
\affiliation{National Institute of Technology Silchar, Silchar 788010, India\\}


\begin{abstract}
 
The di-leptons and di-neutrinos observed in the final states of flavor-changing neutral b decays provide an ideal platform for probing physics
beyond the standard model. Although the latest measurements of $R_{K^{(*)}}$ agree well with the standard model prediction, there exists 
several 
other observables such as  $P_5^{\prime}$, $\mathcal{B}(B_s\to \phi \mu^{+}\mu^{-})$ and $\mathcal{B}(B_s\to \mu^{+}\mu^{-})$ in 
$b\to s \ell\ell$ transition decays that shows deviation from the standard model prediction.
Similalry, very recently Belle II collaboration reported a more precise upper bound of $\mathcal{B}(B\to K^+\nu\bar{\nu}) < 4.1\times 10^{-5}$
by employing a new inclusive tagging approach and it also deviates from the standard model expectation.
The $b\to s l^{+}l^{-}$ and $b\to s\nu\bar{\nu}$ transition decays are related not only in the standard model but also in beyond the standard 
model 
physics due to $SU(2)_L$ gauge symmetry, and can be most effectively investigated using the standard model effective field theory formalism. 
Additionally, the $b\to s\nu\bar{\nu}$ decay channels are theoretically cleaner than the corresponding $b\to s l^{+}l^{-}$ decays, as these
processes do not get contributions from non-factorizable corrections and photonic penguin contributions.
In this context, we study $\Lambda_b\to (\Lambda^*(\to pK^-), \Lambda(\to p\pi))({\mu}^{+}\mu^{-},\,\nu\bar{\nu})$ baryonic decays undergoing 
$b\to s \ell^{+}\ell^{-}$ and $b\to s\nu\bar{\nu}$ quark level transitions in a standard model effective field theory formalism. 
We give predictions of several observables pertaining to these decay channels in the standard model and in case of several new physics 
scenarios.
 
\end{abstract}
 \maketitle
 \section{Introduction}
In high-energy physics experiments, such as those at particle accelerators, it is possible to produce and detect intermediate states of 
quantum particles that have much greater mass than the initial and final particles. These intermediate states are often short-lived and can 
only be observed through the detection of their decay products. In this context, study of flavor changing charged current (FCCC) and neutral 
current (FCNC) transitions of $b$ hadrons is crucial as they can provide important information regarding such intermediate quantum states.
Moreover, FCNC transition decays are, in principle, more sensitive to various new physics~(NP) effects as they proceed either via loop level 
or box level diagrams where the intervention of heavier particles comes into the picture. Hence, study of these decays
would offer a powerful tool to search for NP that lies beyond the standard model~(SM).
Over the past several years, the FCNC B decays have been the center of attention of the particle physics community especially due
to discrepancies observed at BaBar, Belle and more recently at LHCb. The measured values of
the lepton flavor sensitive observable such as the ratio of branching fractions $R_K$ and $R_{K^*}$ in 
$B \to K^{(*)}\, \ell \bar{\ell}$ $(\ell \in e, \mu)$ decays deviate from the SM prediction.
These discrepancies hint for a possible violation of lepton flavor universality (LFU) in $b \to s\,l^+\,l^-$ transition decays. 

Earlier LHCb measurement of $R_K$ in $q^2\in[1.1,6.0]$ $\rm GeV^2$ showed $3.1\sigma$ deviation
from the SM expectation~\cite{LHCb:2021trn}. 
Similalry, earlier measurements of $R_{K^*}$ from both LHCb~\cite{LHCb:2017avl,LHCb:2020lmf} and Belle~\cite{Belle:2019oag} in 
$q^2 \in[0.045,1.1]$ and $q^2\in[1.1,6.0]$ $\rm GeV^2$ bins showed $2.2-2.5\sigma$ deviation~\cite{Bordone:2016gaq,Hiller:2003js} from SM.
However, very recent LHCb results~\cite{LHCb:2022qnv,LHCb:2022zom}, announced in December 2022, has completely changed the
entire scenario. The latest measured values of $R_K =0.994^{+0.090}_{-0.082} (\rm stat) ^{+0.027}_{-0.029} (\rm syst)$
and $R_{K^*} = 0.927^{+0.093}_{-0.087} (\rm stat) ^{+0.034}_{-0.033} (\rm syst)$ in $q^2 \in[0.045,1.1]$ $\rm GeV^2$
and $R_K =0.949^{+0.042}_{-0.041} (\rm stat) ^{+0.023}_{-0.023} (\rm syst)$
and $R_{K^*} = 1.027^{+0.072}_{-0.068} (\rm stat) ^{+0.027}_{-0.027} (\rm syst)$ in $q^2\in[1.1,6.0]$ $\rm GeV^2$
show an overall agreement with the SM prediction with $0.2$ standard deviation~\cite{LHCb:2022qnv,LHCb:2022zom}.

Although $R_K$ and $R_{K^*}$ seem to be SM like, the possibilities of NP cannot be completely ruled out.
Apart from $R_K$ and $R_{K^*}$, there are several other observables where the discrepancy between the measured value and the SM prediction 
still exists. Measurement of $P_5^{\prime}$ from LHCb~\cite{LHCb:2013ghj,LHCb:2015svh} and 
ATLAS~\cite{ATLAS:2018gqc} show $3.3\sigma$ deviation from the SM prediction. Similarly, CMS~\cite{CMS}
and Belle~\cite{Belle:2016xuo} measurements show $1\sigma$ and $2.6\sigma$ deviations, respectively
~\cite{Descotes-Genon:2012isb,Descotes-Genon:2013vna,Descotes-Genon:2014uoa}.
Again, the measured value of the branching fraction of $B_s \to \phi\, \mu \bar{\mu}$
in $q^2\in[1.1,6.0]$ $\rm GeV^2$ deviates from the SM prediction at $3.3\sigma$~\cite{LHCb:2021zwz,LHCb:2013tgx,LHCb:2015wdu}.
Moreover, measurements of the ratio of branching ratio~\cite{LHCb:2021lvy} $R_{K_s^0}$ and $R_{K^{*+}}$, isospin partners of $R_K$ and 
$R_{K^*}$, also deviate from the SM prediction at $1.4\sigma$ and $1.5\sigma$, respectively.

There exists another class of FCNC transition decays with neutral leptons in the final state that are mediated via $b \to s \nu \bar{\nu}$ 
quark level transitions.
Theoretically these di-neutrino channels are clean as they do not suffer from hadronic uncertainties beyond the form factors
such as the non-factorizable corrections and photon penguin contributions. However, they are very challenging from the
experimental point of view due to the presence of neutrinos in the final state.
Inspite of that BaBar, Bell/Belle II has managed to provide the upper bounds of $B \to K^{(*)}\, \nu \bar{\nu}$ decays
to be $\mathcal{B}(B^+ \to K^+\, \nu \bar{\nu}) \le 4.1 \times 10^{-5}$~\cite{Dattola:2021cmw} and
$\mathcal{B}(B^0 \to K^{0*}\, \nu \bar{\nu}) \le 1.8 \times 10^{-5}$, respectively. Combined with the previous measurements from Belle
and BaBar one estimates the world average value of the branching fraction to
be $\mathcal{B}(B^+ \to K^+\, \nu \bar{\nu}) \le (1.1 \pm 04) \times 10^{-5}$~\cite{Dattola:2021cmw}.
A combined analysis of $b \to s \ell \bar{\ell}$ and $b \to s \nu \bar{\nu}$ decays is theoretically well motivated as
these two channels are closely related not only in the SM but also in beyond the SM under $SU(2)_L$ gauge symmetry. Moreover, a more
precise measurements of $B \to K^{(*)}\, \nu \bar{\nu}$ branching fraction in future may provide useful insight into NP that may be
present in $b \to s\,l^+\,l^-$ transition decays.

Various analyses, both model-dependent and model-independent, have been performed to account for these anomalies. A non-exhaustive 
compilation of relevant literature can be found in the references~\cite{Rajeev:2021ntt,Mohapatra:2021ynn,Li:2011nf,Huang:2018rys,
Falahati:2014yba, Ahmed:2010tt,Capdevila:2017bsm,Bashiry:2009wq,Wang:2012ab,Li:2010ra,Rajeev:2020aut,
 Browder:2021hbl,Bobeth:2001jm,Altmannshofer:2009ma,Descotes-Genon:2020buf, Fajfer:2018bfj}.
To confirm the presence of NP, we need to perform measurements of similar observables in different decay
processes that proceed via same quark level transitions.
Similarly, it is very important to perform a detailed angular analysis in order to look for several form factor independent angular
observables which are sensitive to NP. 
In this context, baryonic $\Lambda_b^0 \to \Lambda^{(*)}\,(\to pK)\,\mu^+\mu^-$ decay mode has got lot of attention. The recent measurement 
from LHCb suggests that although the ratio $R_{pK}$ is compatible with SM, there is suppression in 
$\mathcal{B}(\Lambda_b \to pK \mu \bar{\mu})$ compared to $\mathcal{B}(\Lambda_b \to pK e \bar{e})$~\cite{LHCb:2019efc}.
To interprit this result, it is essential to have a precise theoretical knowledge of various excited states of $\Lambda$ baryon
contributing to $pK$ region.  
The $\Lb$ decay to $\Lst\equiv \Lst(1520)$ has the largest contribution among the various semileptonic modes of $\Lb$ decays to hadrons. 
Due to its spin parity of $J^P=3/2^-$ and strong decay into the $N\bar{K}$ pair, the $\Lst$ is readily distinguishable from nearby hadrons, 
including the $\Lambda(1600)$, $\Lambda(1405)$, and weakly decaying $\Lambda(1116)$, which have a spin parity of $J^P=1/2^\pm$.
In Ref~\cite{Meinel:2020owd,Meinel:2021mdj}, the authors calculate the LQCD form factors
in the weak transition of $\Lambda_b\to\Lambda(1520)$ decay, while in Ref.~\cite{Descotes-Genon:2019dbw} and Ref.~\cite{Das:2020cpv}, the 
authors performed angular analyses of $\Lambda_b\to\Lambda\ell^+\ell^-$ decays for massless and massive leptons, respectively. Additionally, 
in Ref~\cite{Amhis:2020phx}, the authors investigated the angular distributions of $\Lambda_b\to\Lambda(1520)\ell^+\ell^-$ and discussed the 
potential for identifying NP effects. Similalry, the authors in Ref\cite{Li:2022nim} study the 
$\Lambda_b\to\Lambda(1520)(\to N\bar{K})\ell^+\ell^-$ process with $N\bar{K}=\{pK^-,n\bar{K}^0\}$ and examine several angular observables. 
The study is performed with a set of operators where the SM operator basis is supplemented with its chirality flipped counterparts
and new scalar and pseudoscalar operators. The three-body light-front quark model based on the gaussian expansion method is used to 
systematically investigate the $\Lambda_b\to\Lambda(1520)(\to N\bar{K})\ell^+\ell^-$ ($\ell=e,\mu,\tau$) decay process.
Several theoretical methods, such as lattice QCD (LQCD)~\cite{Detmold:2012vy,Detmold:2016pkz}, QCD sum rules (QCDSR) \cite{Chen:2001sj},
light-cone sum rule (LCSR)~\cite{Aslam:2008hp,Wang:2008sm,Wang:2009hra,
Aliev:2010uy,Wang:2015ndk}, covariant quark model (CQM)~\cite{Gutsche:2013pp}, nonrelativistic quark model~\cite{Mott:2011cx}, 
and Bethe-Salpeter approach~\cite{Liu:2019igt}, have been used to study the rare decay $\Lambda_b\to\Lambda\ell^+\ell^-$.
The initial measurement of the decay was conducted by the CDF Collaboration~\cite{CDF:2011buy}, followed by a subsequent measurement by the 
LHCb Collaboration~\cite{LHCb:2015tgy,LHCb:2018jna}. 
In ref~\cite{Huang:1998ek} QCD sum rules were used to calculate the $\Lambda_b\to \Lambda$ transition form factors and to study the 
unpolarized decay. The form factors for 
$\Lambda_b\to\Lambda$ at large recoil were analyzed using a sum-rule approach to study spectator-scattering corrections~\cite{Feldmann:2011xf}.
Light-cone distribution amplitude of $\Lambda_b$ wave function was studied in \cite{Ali:2012pn,
Bell:2013tfa,Braun:2014npa} to further understand the theoretical aspects. A model-independent analysis for unpolarized 
$\Lambda_b\to\Lambda(\to N\pi)\ell^+\ell^-$ decay was performed in \cite{Das:2018sms, Das:2018iap, Das:2020cpv, Roy:2017dum,Yan:2019tgn} 
using a complete set of dimension-six operators. 
The angular distribution of the decay with unpolarised $\Lambda_b$ baryon has been explored
in Refs.\cite{Boer:2014kda,Yan:2019tgn}, while in Ref.\cite{Blake:2017une}, the study involved polarized $\Lambda_b$ baryon. Furthermore,
in Ref.\cite{Blake:2019guk}, the $b\to s\mu^+\mu^-$ Wilson coefficients were examined by utilizing the complete angular distribution of the 
rare decay $\Lambda_b\to\Lambda(\to p\pi)\mu^{+}\mu^{-}$ measured by the LHCb Collaboration~\cite{LHCb:2018jna}. 
Similalry, in Ref\cite{Chen:2000mr}, the authors calculate the branching fraction of 
$\Lambda_b\to\Lambda\nu\bar{\nu}$ decay by taking the polarised $\Lambda_b$ and $\Lambda$. Moreover, in Ref~\cite{Sirvanli:2007yq}, the 
authors analyse $\Lambda_b\to\Lambda\nu\bar{\nu}$ decay
by considering the $Z^{'}$ model. Here the authors calculate the branching ratio as well as the longitudinal, transversal and normal 
polarizations of the di-neutrino decay channel of the baryonic decay $\Lambda_b\to\Lambda$ within the SM as well as in the presence of 
leptophobic $Z^{'}$ model.

In this paper, we study the implication of $b \to s\,l^+\,l^-$ anomalies on 
$\Lambda_b \to \Lambda^{(*)}\,l^+\,l^-$ and 
$\Lambda_b \to \Lambda^{(*)}\,\nu \bar{\nu}$ decays in a model independent way. Our work differs significantly from others. For NP analysis, we
construct several $1D$ and $2D$ NP scenarios emerging out of dimension six operators in the standard model effective field theory~(SMEFT) 
formalism. We obtain the allowed NP parameter space by performing a global fit to the $b \to s\,l^+\,l^-$ data. Moreover, we also use the
measured upper bound on $\mathcal{B}(B \to K^{(*)}\nu\bar{\nu})$ to check the compatibility of our fit results.

The paper is organized as follows. In Section~\ref{theory}, we start with a brief description of the SMEFT framework 
and write down the effective Hamiltonian for the $b\to s\nu\bar{\nu}$ and $b\to s\,l^+\,l^-$ quark level transition decays.
Subsequently, we report all the relevant formulae for the observables in Section~\ref{theory}. 
In Section~\ref{result}, we first report all the input parameters that are used for our analysis. A detailed discussion of the results
pertaining to 
$\Lambda_b\to (\Lambda^*(\to pK^-), \Lambda(\to p\pi)){\mu}^{+}\mu^{-}$ and 
$\Lambda_b\to (\Lambda^*(\to pK^-), \Lambda(\to p\pi))\nu\bar{\nu}$ baryonic
decay observables in the SM and in case of NP scenarios are also presented. Finally, we conclude with a brief summary of our results in 
Section~\ref{conc}.

\section{Theory}
\label{theory}

Till date, no direct evidence of new particles near the electroweak scale has been observed from searches conducted in the Large Hadron 
Collider~(LHC). Nevertheless, these searches provide indirect evidence supporting the existence of NP at a scale beyond the 
electroweak scale. To explore indirect signatures of NP in a model-independent way, the SMEFT 
framework offers a more efficient approach.
The SMEFT Lagrangian explains particle interactions in the SM and in all possibleextensions of SM. 
It is constructed by incorporating higher-dimensional operators into the SM Lagrangian while maintaining the 
${SU(3)}_C \times {SU(2)}_L \times {U(1)}_Y$ gauge symmetry.
These higher-dimensional operators are suppressed by a factor that depends on a new energy scale. The SMEFT Lagrangian comprises all sets of 
these higher-dimensional operators that are consistent with the underlying gauge symmetry. For investigating NP beyond the SM 
at low energies, this framework provides an excellent platform. From the fundamental aspect of the electroweak theory, the left-handed 
charged leptons are related to neutral leptons through the $SU(2)_L$ symmetry. In this study, we concentrate on the connection between 
$b\to s\,l^+\,l^- $ and $b\to s \nu\bar{\nu}$ transition decays within the SMEFT framework by considering dimension six operators.  
If no new particles are observed at the LHC, it will imply a NP scale that is greater than the energy scale of the LHC. The SMEFT 
analysis would be crucial in this situation as it offers a way to examine the implications of NP indirectly by evaluating their 
effects on SM low energy processes.

The effective Lagrangian corresponding to dimension six operators is expressed as~\cite{Grzadkowski:2010es}
\begin{equation}
 \mathcal{L}^{(6)}=\sum_i \frac{c_i}{\Lambda^2}\, \mathcal{Q}_i\,.
\end{equation}
Among all the operators, the relevant operators contributing to both $b \to s\nu\bar{\nu}$ and 
$b \to s\,l^+\,l^-$ decays are 
\begin{eqnarray}\label{smeft-o}
 \mathcal{Q}_{Hq}^{(1)}=i(\bar{q}_L \gamma_{\mu} q_L)H^\dagger D^{\mu} H, \hspace{0.5cm} 
 \mathcal{Q}_{Hq}^{(3)}=i(\bar{q}_L \gamma_{\mu} \tau^a q_L)H^\dagger D^{\mu} \tau_a H, \hspace{0.5cm}
 \mathcal{Q}_{Hd}=i(\bar{d}_R \gamma_{\mu} d_R)H^\dagger D^{\mu} H, \nonumber \\
 \mathcal{Q}_{ql}^{(1)}=(\bar{q}_L \gamma_{\mu} q_L)(\bar{l}_L \gamma^{\mu} l_L), \hspace{0.5cm}
 \mathcal{Q}_{ql}^{(3)}=(\bar{q}_L \gamma_{\mu} \tau^a q_L)(\bar{l}_L \gamma^{\mu} \tau_a l_L), \hspace{0.5cm}
 \mathcal{Q}_{dl}=(\bar{d}_R \gamma_{\mu} d_R)(\bar{l}_L \gamma^{\mu} l_L)\,.
 \end{eqnarray}
Similarly, the operators contributing only to $b \to s\,l^+\,l^-$ decays are 
 \begin{eqnarray}\label{smeft-o2}
 \mathcal{Q}_{de}=(\bar{d}_R \gamma_{\mu} d_R)(\bar{e}_R \gamma^{\mu} e_R), \hspace{0.5cm}
 \mathcal{Q}_{qe}=(\bar{q}_L \gamma_{\mu} q_L)(\bar{e}_R \gamma^{\mu} e_R)\,.
\end{eqnarray}
At low energy, the most general $\Delta F=1$ effective Hamiltonian governing both  $b \to s\nu\bar{\nu}$ 
and $b \to s\,l^+\,l^-$ decays can be written as~\cite{Altmannshofer:2009ma,Buras:2014fpa},
\begin{equation}
 \mathcal{H}_{eff}=-\frac{4 G_F}{\sqrt{2}}\, V_{tb} V_{ts}^{*}\, \frac{e^2}{16\pi^2}\, \sum_i C_i\, \mathcal{O}_i\, + h.c.,
\end{equation}
where $G_F$ is the Fermi coupling constant, $|V_{tb} V_{ts}^{*}|$ are the associated Cabibbo Kobayashi Maskawa~(CKM) matrix elements.
The sum $i=L,R$ comprises the operators $\mathcal{O}_{L,R}$ with the corresponding WCs $C_{L,R}$ contributing to 
$b \to s\nu\bar{\nu}$ decays. They are
\begin{equation}
\mathcal{O}_L=(\bar{s}\gamma_{\mu}P_L b)(\bar{\nu}\gamma^{\mu}(1-\gamma_5)\nu), \hspace{0.5cm}
\mathcal{O}_R=(\bar{s}\gamma_{\mu}P_R b)(\bar{\nu}\gamma^{\mu}(1-\gamma_5)\nu).
\end{equation}
Here, $P_{L,R}=(1\mp\gamma_5)/2$ represents the projection operator. 
In the SM, $C_R^{\rm SM} = 0$ and the value of $C_L^{\rm SM}$ is calculated to be
\begin{equation}
 C_{L}^{\rm SM}=-X_t/s_{w}^2=-6.38 \pm 0.06, \hspace{1cm} X_t=1.469 \pm 0.017, \hspace{1cm} s_{w}^2=0.23126(5).
\end{equation}
Similarly, for $i=9^{(\prime)},10^{(\prime)}$, the sum comprises the operators $\mathcal{O}_{9^{(\prime)},10^{(\prime)}}$ with the 
corresponding WCs $C_{9^{(\prime)},10^{(\prime)}}$ that contribute to $b \to s\,l^+\,l^-$ decays. The operators are
\begin{equation}
 \mathcal{O}_9^{(\prime)}=(\bar{s}\gamma_{\mu}P_{L(R)} b)(\bar{l}\gamma^{\mu}l), \hspace{0.5cm}
\mathcal{O}_{10}^{(\prime)}=(\bar{s}\gamma_{\mu}P_{L(R)} b)(\bar{l}\gamma^{\mu}\gamma_5 l).
\end{equation}

In the presence of dimension six SMEFT operators, the WCs $C_{9,10,L}$ and $C_{9^{\prime},10^{\prime},R}$ get modified. They can be expressed 
as follows.~\cite{Buras:2014fpa}
 \begin{eqnarray}
  C_9&=&C_9^{\rm SM}\, + \widetilde{c}_{qe}\, + \widetilde{c}_{ql}^{(1)}\, + \widetilde{c}_{ql}^{(3)}\, - \zeta \widetilde{c}_{Z} \nonumber \\
  C_{10}&=&C_{10}^{\rm SM}\, + \widetilde{c}_{qe}\, - \widetilde{c}_{ql}^{(1)}\, - \widetilde{c}_{ql}^{(3)}\, + \widetilde{c}_{Z} \nonumber \\
  {C}_{L}^{\nu}&=&C_{L}^{\rm SM}\, + \widetilde{c}_{ql}^{(1)}\, - \widetilde{c}_{ql}^{(3)}\, + \widetilde{c}_{Z} \nonumber \\
  C_{9}^{\prime}&=& \widetilde{c}_{de}\, + \widetilde{c}_{dl}\, - \zeta \widetilde{c}_{Z}^{\prime} \nonumber \\
  C_{10}^{\prime}&=& \widetilde{c}_{de}\, - \widetilde{c}_{dl}\, + \widetilde{c}_{Z}^{\prime} \nonumber \\
  {C}_{R}^{\nu}&=& \widetilde{c}_{dl}\, + \widetilde{c}_{Z}^{\prime}\,,
 \end{eqnarray}
where, $\widetilde{c}_{Z}=\frac{1}{2}(\widetilde{c}_{Hq}^{(1)}+\widetilde{c}_{Hq}^{(3)})$, $\widetilde{c}_{Z}^{\prime}=
\frac{1}{2}(\widetilde{c}_{Hd})$ and $\zeta \approx 0.08$ represents the small vector coupling to charged leptons.

\subsection{Differential decay distribution and $q^2$ dependent observables for 
$\Lb\to \Lst(\to pK^-)\ell^+\ell^-$ decays}
The four-fold angular distribution for $\Lb\to \Lst( \to pK^-)\ell^+\ell^-$ decay can be expressed as~\cite{Das:2020cpv}
 \begin{eqnarray}\label{eq:fourfold}
\frac{d^4\mathcal{B}}{dq^2 d\cl d\cLst d\phi} &=& \frac{3}{8\pi} \bigg[\bigg(K_{1c}\cl + K_{1cc}\ccl + K_{1ss}\ssl \bigg)\ccLst\, \nn\\ 
&&~~~~+\bigg(K_{2c}\cl + K_{2cc}\ccl + K_{2ss}\ssl  \bigg)\ssLst\, \nn\\
&&~~~~+\bigg(K_{3ss}\ssl \bigg)\ssLst\cos\phi + \bigg(K_{4ss}\ssl\bigg)\ssLst\sin\phi\cos\phi \,\nn\\
&&~~~~+\bigg(K_{5s}\sl + K_{5sc}\sl\cl \bigg)\sLst\cLst\cos\phi\, \nn\\
&&~~~~+\bigg(K_{6s}\sl + K_{6sc}\sl\cl \bigg)\sLst\cLst\sin\phi\,
\bigg] \,, 
\end{eqnarray}
where $\theta_{\Lambda^{*}}$ represents the angle formed by the proton with the daughter baryon $\Lambda^{*}$ in the rest frame of
$\Lambda_b$. Similarly, in the rest frame of the lepton pair, $\theta_\ell$ denotes the angle formed by the $\ell^-$ with respect to the
direction of the daughter baryon $\Lambda^{*}$. Moreover, in the rest frame of $\Lambda_b$, $\phi$ defines the angle between the planes 
containing $p\,K^-$ and the lepton pair. 
The angular coefficients $K_{\{\cdots \}}$, ${\{\cdots \}}=1c, \cdots 6sc$, can be expressed as
\begin{equation}
K_{\{\cdots \}} = {K}_{\{\cdots \}} + \frac{m_\ell}{\sqrt{q^2}} {K}^\prime_{\{\cdots \}} + \frac{m_\ell^2}{q^2} 
{K}^{\prime\prime}_{\{\cdots \}}\,.
\end{equation}
Here the first term ${K}$ corresponds to massless leptons, whereas, ${K}^{\prime}$ and ${K}^{\prime\prime}$ correspond to 
linear~($\mathcal{O}(m_\ell/\sqrt{q^2})$) and 
quadratic~($\mathcal{O}(m_\ell^2/q^2)$) mass 
corrections, respectively.
The explicit expressions for $K_{\{\cdots \}}$, ${K}^\prime_{\{\cdots \}}$ and ${K}^{\prime\prime}_{\{\cdots \}}$ in terms of 
transversely amplitude are taken Ref~\cite{Das:2020cpv}.

From the Differential decay distributions, one can construct several physical observables. 
\begin{itemize}
\item The differential branching ratio $d\mathcal{B}/dq^2$, the lepton forward-backward asymmetry $A_{FB}^l (q^2)$, 
the fraction of longitudinal polarization $F_L (q^2)$ and the ratio of branching fraction $R_{\Lambda^*}(q^2)$ are defined as 
\begin{eqnarray}
\frac{d\mathcal{B}}{dq^2} &=& \frac{1}{3}\bigg[K_{1cc} + 2K_{1ss} + 2K_{2cc} + 4K_{2ss} + 2K_{3ss}  \bigg]\, 
\hspace{0.5cm}
F_L = 1 - \frac{2(K_{1cc} + 2K_{2cc} )}{K_{1cc} + 2(K_{1ss} + K_{2cc} + 2K_{2ss} + K_{3ss} ) }\\ \nonumber
A^\ell_{\rm FB} &=& \frac{3(K_{1c} + 2K_{2c} )}{2\big[ K_{1cc} + 2(K_{1ss} + K_{2cc} + 2K_{2ss} + K_{3ss} ) \big] } \hspace{1cm}  R_{\Lambda^*}(q^2) = \frac{d\mathcal{B}/dq^2|_{\mu-mode}}{d\mathcal{B}/dq^2|_{e-mode}}
\end{eqnarray}

\item We define several angular observables such as $\hat{K}_{1c}$, $\hat{K}_{1cc}$,
$\hat{K}_{1ss}$, $\hat{K}_{2c}$, $\hat{K}_{2cc}$, $\hat{K}_{2ss}$, $\hat{K}_{3ss}$, $\hat{K}_{4ss}$,
, $\hat{K}_{4s}$, $\hat{K}_{5s}$. They are 
\begin{eqnarray}
 \hat{K}_{1c}&=&\frac{{K}_{1c}}{{d\mathcal{B}}/{dq^2}} \hspace{0.5cm} 
  \hat{K}_{1cc}=\frac{{K}_{1cc}}{{d\mathcal{B}}/{dq^2}}  \hspace{0.5cm} 
   \hat{K}_{1ss}=\frac{{K}_{1ss}}{{d\mathcal{B}}/{dq^2}}\hspace{0.5cm} 
   \hat{K}_{2c}=\frac{{K}_{2c}}{{d\mathcal{B}}/{dq^2}} \hspace{0.5cm} 
   \hat{K}_{2cc}=\frac{{K}_{2cc}}{{d\mathcal{B}}/{dq^2}}\hspace{0.5cm} \nonumber \\ 
    \hat{K}_{2ss}&=&\frac{{K}_{2ss}}{{d\mathcal{B}}/{dq^2}}\hspace{0.5cm}
     \hat{K}_{3ss}=\frac{{K}_{3ss}}{{d\mathcal{B}}/{dq^2}} \hspace{0.5cm} 
      \hat{K}_{4ss}=\frac{{K}_{4ss}}{{d\mathcal{B}}/{dq^2}} \hspace{0.5cm}
       \hat{K}_{5s}=\frac{{K}_{5s}}{{d\mathcal{B}}/{dq^2}}\,. \hspace{0.5cm}
\end{eqnarray}
\end{itemize}

It is important to note that the angular coefficients ($K_{1c}$, $K_{2c}$), ($K_{1cc}$, $K_{2cc}$), and ($K_{1ss}$, $K_{2ss}$) exhibit a 
strict relation in the SM. That is
\begin{equation}
 \frac{K_{1c}}{K_{2c}} = 4\, ,\quad \frac{K_{1cc}}{K_{2cc}}=4\, ,\quad \frac{K_{1ss}}{K_{2ss}} = 4\, .
\end{equation}

\subsection{Differential decay distribution and $q^2$ dependent observables for $\Lambda_b\to\Lambda(\to p \pi)\ell^+\ell^-$ }
The four fold angular distribution for $\Lambda_b\to\Lambda(\to p\pi)\ell^+\ell^-$ is defined as~\cite{Das:2018iap}
\begin{eqnarray}
\frac{d^4\mathcal{B}}{dq^2d\cos\theta_\ell d\cos\theta_\Lambda d\phi} &=& \frac{3}{8\pi}
(K_{1ss} \sin^2\theta_\ell + K_{1cc} \cos^2\theta_\ell + K_{1c} \cos\theta_\ell )\, \nn\\
&+& (K_{2ss} \sin^2\theta_\ell + K_{2cc} \cos^2\theta_\ell + K_{2c} \cos\theta_\ell ) \cos\theta_\Lambda \, \nn\\
&+&( K_{3sc}\sin\theta_\ell\cos\theta_\ell + K_{3s}\sin\theta_\ell ) \sin\theta_\Lambda \sin\phi \, \nn\\
&+& ( K_{4sc}\sin\theta_\ell\cos\theta_\ell + K_{3s}\sin\theta_\ell ) \sin\theta_\Lambda \cos\phi \, .
\end{eqnarray}
where the angular coefficients $K_{ijk}$ can be expressed as
\begin{equation}
 K_{ijk} = {K}_{{ijk}} + \frac{m_\ell}{\sqrt{q^2}} {K}_{{ijk}}^\prime + \frac{m_\ell^2}{q^2}
{K}_{{ijk}}^{\prime\prime}\,,
\end{equation}
with ${ijk} = 1ss \cdots\ 4s$. The explicit expressions for ${K}$, ${K}^{\prime}$ and ${K}^{\prime\prime}$ are taken
from Ref~\cite{Das:2018iap}.

We define several physical observables pertaining to this decay mode.
\begin{itemize}

\item Differential branching ratio $d\mathcal{B}/dq^2$, the lepton forward-backward asymmetry $A_{FB}^l (q^2)$, 
the fraction of longitudinal polarization $F_L (q^2)$ and the ratio of branching fraction $R_{\Lambda}(q^2)$ are defined as
\begin{eqnarray}
	\frac{d\mathcal{B}}{dq^2} = 2K_{1ss} + K_{1cc}\, \hspace{0.5cm}
	F_L = \frac{ 2K_{1ss} - K_{1cc} }{ 2K_{1ss} + K_{1cc} }\, ,\quad 
	A^\ell_{\rm FB} = \frac{3}{2} \frac{K_{1c}}{ 2K_{1ss} + K_{1cc} }\, ,\quad\quad 
        R_{\Lambda}(q^2) = \frac{d\mathcal{B}/dq^2|_{\mu-mode}}{d\mathcal{B}/dq^2|_{e-mode}}
\end{eqnarray} 
\item  Angular observables such as $\hat{K}_{1c}$, $\hat{K}_{1cc}$,
$\hat{K}_{1ss}$, $\hat{K}_{2c}$, $\hat{K}_{2cc}$, $\hat{K}_{2ss}$, $\hat{K}_{3ss}$, $\hat{K}_{3sc}$,
$\hat{K}_{4sc}$, $\hat{K}_{4s}$ are defined as
\begin{eqnarray}
 \hat{K}_{1c}&=&\frac{{K}_{1c}}{{d\mathcal{B}}/{dq^2}} \hspace{0.5cm} 
  \hat{K}_{1cc}=\frac{{K}_{1cc}}{{d\mathcal{B}}/{dq^2}}  \hspace{0.5cm} 
   \hat{K}_{1ss}=\frac{{K}_{1ss}}{{d\mathcal{B}}/{dq^2}}\hspace{0.5cm} 
   \hat{K}_{2c}=\frac{{K}_{2c}}{{d\mathcal{B}}/{dq^2}} \hspace{0.5cm} 
   \hat{K}_{2cc}=\frac{{K}_{2cc}}{{d\mathcal{B}}/{dq^2}}\hspace{0.5cm} \nonumber \\ 
    \hat{K}_{2ss}&=&\frac{{K}_{2ss}}{{d\mathcal{B}}/{dq^2}} \hspace{0.5cm}
     \hat{K}_{3sc}=\frac{{K}_{3sc}}{{d\mathcal{B}}/{dq^2}} \hspace{0.5cm} 
      \hat{K}_{3s}=\frac{{K}_{3s}}{{d\mathcal{B}}/{dq^2}} \hspace{0.5cm} 
      \hat{K}_{4sc}=\frac{{K}_{4sc}}{{d\mathcal{B}}/{dq^2}} \hspace{0.5cm}
       \hat{K}_{4s}=\frac{{K}_{4s}}{{d\mathcal{B}}/{dq^2}} \hspace{0.5cm}
\end{eqnarray}

\end{itemize}

\section{Results}
\label{result}
\subsection{Input parameters}
The numerical values of all the input parameters used in the paper are summarized in the Table~\ref{tab_input}.
Input parameters, such as the masses of mesons and quarks are expressed in GeV units, the Fermi coupling constant $G_F$ is in GeV$^{-2}$
units and the life time of $\Lambda_b$ baryon is in seconds. 
\begin{table}[htbp]
\centering
\scalebox{0.9}{
\begin{tabular}{|cc|cc|cc|cc|cc|}
\hline
Parameter & Value & Parameter & Value & Parameter & Value & Parameter & Value & Parameter & Value \\
\hline
\hline
$m_e$ & 0.000511 &$m_{\mu}$ & 0.105658 & $ \alpha_e(m_b) $  & $1/127.925(16)$ & $|V_{tb}V_{ts}^\ast|$  & $0.0401 \pm 0.0010$  & $m_c(\overline{\text{MS}})$ & 1.28 GeV \\
$\mLb$ & 5.619 GeV  & $\mu_b$ & $4.8$ & $\mL$ & $1.115$ & $\tau_{\Lambda_b}$ & $(1.470\pm 0.010)\times 10^{-12}$
& $\mLb$ & 5.619 GeV  \\
$m_b^{\bar{MS}}$ & 4.2 & $m_c^{\bar{MS}}$ & 1.28 & $m_b^{pole}$ & 4.8 &  $\mL$ & $1.115$  & $\mathcal{B}_\Lst$ & $0.45\pm0.01$\\
$\mathcal{B}_\Lambda$ & $0.642\pm0.013$  & $\alpha_{\Lambda}$ & 0.443 & $\alpha_{\Lambda^{'}}$  & 0.333 & 
&  & & \\
\hline
\hline
\end{tabular}}
\caption{Input parameters~\cite{ParticleDataGroup:2022pth, Das:2018iap, Das:2020cpv}.}
\label{tab_input}
\end{table}

For hadronic inputs such as $\Lambda_b \to \Lambda^* $ form factors, we use the values reported in Ref.~\cite{Mott:2011cx}, and for 
$\Lambda_b \to \Lambda $ form factors, we use the LQCD results of Ref~\cite{Detmold:2016pkz}.
The relevant formula for the $\Lb\to \Lst$ form factors pertinent for our discussion is as follows~\cite{Mott:2011cx}
\begin{equation}
F(\hat{s})=(a_0+a_2 p_{\Lambda}^2+a_4p_{\Lambda}^4)\exp\Big{(}-\frac{3m_q^2}{2\tilde{m}_{\Lambda}^2} \frac{p_{\Lambda}^2}
{\alpha_{\Lambda\Lambda^{'}}^2}\Big{)} \,,
\end{equation}
where 
\begin{eqnarray}
p_{\Lambda}=\frac{ {m_{\Lambda}}_b }{2}\sqrt{\phi(\hat{s})}\,, \qquad\qquad
\phi{(\hat{s})}= {(1-r)}^2 - 2(1+r)\hat{s}+\hat{s}^2\,, \qquad\qquad
\alpha_{\Lambda\Lambda^{'}}=\sqrt{\frac{\alpha_{\Lambda}^2+{\alpha^2_{\Lambda^{'}}}}{2}}\,. 
\end{eqnarray}
Here $r={m^2_{\Lambda^*}}/{m^2_{{\Lambda}_b}}$ and  
$\hat{s}\equiv q^2/{{m_{\Lambda}}^2_b}$. We consider $5\%$ uncertainty in the input parameters 
 $F_i\in (i= 1...4)$, $G_i\in (i= 1...4)$ and $H_i(i= 1...6)$. The values 
 of these parameter, taken from Ref~\cite{Mott:2011cx}, are reported in Table~\ref{ff_lstr}. 
 
 \begin{table}[htbp]
\centering
\resizebox{\columnwidth}{!}{
\begin{tabular}{|c|c|c|c|c|c|c|c|c|c|c|c|c|c|c|}
\hline
  &\multicolumn{13}{c}{$\Lambda_b \to \Lambda^* $ form factor inputs}  & \\
 \hline
 $\Lambda^*(1520)$ &  $F_1$  & $F_2$ &  $F_3$& $F_4$ &  $G_1$  & $G_2$ &  $G_3$& $G_4$
 &  $H_1$  & $H_2$ &  $H_3$& $H_4$ & $H_5$ & $H_6$\\
\hline
$a_0$  &-1.66   & 0.544    & 0.126     & -0.0330  & -0.964  &  0.625 &  -0.183     & 0.0530 & 
 -1.08 &  -0.507   & 0.187  & 0.0772 & -0.0517 & 0.0206\\
$a_2$  & -0.295  & 0.194    & 0.00799   & -0.00977 & -0.100  &  0.219 & -0.0380   & 0.0161 &
  -0.0732 & -0.246  & 0.0295  & 0.0267 & -0.0173 & 0.00679\\
$a_4$  & 0.00924 & -0.00420 & -0.000365 & 0.00211  & 0.00264 &  -0.00508  & 0.00351 & -0.00221 &
 0.00464 &  0.00309 & -0.00107  & -0.00217 & 0.00259 & -0.000220\\
 \hline 
\end{tabular}}
 \caption{$\Lambda_b \to \Lambda^*$ form factor inputs~\cite{Mott:2011cx}}
\label{ff_lstr}
\end{table}

Similarly, for $\Lambda_b \to \Lambda$ transition form factors, we use the relevant form factor formula from Ref~\cite{Detmold:2016pkz}.
That is
 \begin{equation}\label{eq:znominal}
f(q^2) = \frac{1}{1-q^2/(m^{f}_{\rm pole})^2} \big[ a^{f}_0 + a_1^{f} z(q^2,t_+) \big]\,.
\end{equation}
To calculate the statistical uncertainties of the observable, we utilize the parameters from the "nominal" fit. However, to estimate the 
systematic uncertainties, we use a "higher-order" fit where the fit function is given by
\begin{equation}\label{eq:zhigher}
f(q^2) = \frac{1}{1-q^2/(m^{f}_{\rm pole})^2} \big[ a^{f}_0 + a_1^{f} z(q^2,t_+) + a_2^{f} (z(q^2,t_+))^2 \big]\,.
\end{equation}
The function $z(q^2,t_+)$ is defined as
\begin{equation}
	z(q^2,t_+) = \frac{ \sqrt{t_+-q^2} - \sqrt{t_+-t_0}  }{ \sqrt{t_+-q^2} + \sqrt{t_+-t_0} }\, ,
\end{equation}
where $t_0 = (\mLb-m_{\Lambda})^2$ and $t_+=(m_B+m_K)^2$. The fit parameters and masses used in our analysis are taken from 
Ref.~\cite{Detmold:2016pkz}. For completeness, we report them in Table~\ref{tab:nominal}. 

\begin{table}
 \setlength{\tabcolsep}{4pt} 
 \renewcommand{\arraystretch}{1.0} 
\begin{tabular}{|c|c|cc|c|}
\hline \hline
 Parameter         &  Value  & \hspace{2ex} &  Parameter         & Value  \\
\hline
$a_0^{f_+}$ & $ 0.4221\pm 0.0188$ &&  $a_1^{g_0}$ & $-1.0290\pm 0.1614$ \\  
 $a_1^{f_+}$ & $-1.1386\pm 0.1683$ &&  $a_1^{g_\perp}$ & $-1.1357\pm 0.1911$ \\  
 $a_0^{f_0}$ & $  0.3725\pm 0.0213$ &&  $a_0^{h_+}$ & $  0.4960\pm 0.0258$ \\  
 $a_1^{f_0}$ & $-0.9389\pm 0.2250$ &&  $a_1^{h_+}$ & $-1.1275\pm 0.2537$ \\  
 $a_0^{f_\perp}$ & $  0.5182\pm 0.0251$ &&  $a_0^{h_\perp}$ & $  0.3876\pm 0.0172$ \\  
 $a_1^{f_\perp}$ & $-1.3495\pm 0.2413$ &&  $a_1^{h_\perp}$ & $-0.9623\pm 0.1550$ \\  
 $a_0^{g_\perp,g_+}$ & $  0.3563\pm 0.0142$ &&  $a_0^{\widetilde{h}_{\perp},\widetilde{h}_+}$ & $  0.3403\pm 0.0133$ \\  
 $a_1^{g_+}$ & $-1.0612\pm 0.1678$ &&  $a_1^{\widetilde{h}_+}$ & $-0.7697\pm 0.1612$ \\  
 $a_0^{g_0}$ & $  0.4028\pm 0.0182$ &&  $a_1^{\widetilde{h}_\perp}$ & $-0.8008\pm 0.1537$ \\  
 \hline 
 $m_{pole}^{f_+}$, $m_{pole}^{f_\perp}$, $m_{pole}^{h_+}$, $m_{pole}^{h_\perp}$ & $5.416$ && $m_{pole}^{f_0}$
 & $5.711$ \\
 $m_{pole}^{g_+}$, $m_{pole}^{g_\perp}$, $m_{pole}^{\widetilde{h}_+}$, $m_{pole}^{\widetilde{h}_\perp}$& $5.750$ && $m_{g_0}$ & $5.367$  \\
\hline
\end{tabular}
 \caption{\label{tab:nominal}$\Lambda_b\to\Lambda$ form factor inputs~\cite{Detmold:2016pkz}.}
\end{table}

\subsection{SM predictions}

In this section, we present the central values and the $1\sigma$ uncertainties of several observables for the
$\Lb\to \Lst(\to pK^-)\ell^+\ell^-$ and $\Lambda_b\to\Lambda(\to p\pi)\ell^+\ell^-$ decay channels. More specifically, we give prediction
of the branching ratio ($BR$), the ratio of branching ratios ($R_{\Lambda^{(*)}}$), the forward-backward asymmetry ($A_{FB}^l$), the 
longitudinal 
polarization fraction ($F_L$) for the $\mu^+\mu^-$ modes, respectively. We also report 
various angular observables such as $\hat{K}_{1ss}$, $\hat{K}_{2c}$, $\hat{K}_{2cc}$, $\hat{K}_{2ss}$, $\hat{K}_{3ss}$, $\hat{K}_{4ss}$, 
$\hat{K}_{4s}$, $\hat{K}_{5s}$ for $\Lb\to \Lst(\to pK^-)\ell^+\ell^-$ decay mode. Similarly, we report angular observables such as 
$\hat{K}_{1c}$, $\hat{K}_{1cc}$, $\hat{K}_{1ss}$, $\hat{K}_{2c}$, $\hat{K}_{2cc}$, $\hat{K}_{2ss}$, $\hat{K}_{3ss}$, $\hat{K}_{3sc}$, 
$\hat{K}_{4sc}$, $\hat{K}_{4s}$ for $\Lambda_b\to\Lambda(\to p\pi)\ell^+\ell^-$ decay mode as well. Moreover, we give predictions of
several observables pertaining to $\Lb\to \Lst(\to pK^-)\nu \bar\nu$ and $\Lambda_b\to\Lambda(\to p\pi)\nu\bar\nu$ decay modes. 
The central values of the observables are obtained using the central values of the input parameters, whereas the uncertainties in each 
observable are determined by varying the uncertainties associated with inputs such as form factors and the CKM matrix elements within 
$1\sigma$ of their central values. For the $\mu^+\mu^-$ final states,
we explore two $q^2$ bins, namely $(1.1 - 6.0)$ and $(14.2 - q^2_{max})$ for the $\Lb\to \Lst(\to pK^-)\ell^+\ell^-$ decay mode and 
$(1.1 - 6.0)$ and $(15.0 - q^2_{max})$ for the $\Lambda_b\to\Lambda(\to p\pi)\ell^+\ell^-$ decay mode, respectively. 
All the results are listed in Table~\ref{sm_rslt_lst} and Table.~\ref{sm_rslt_lst_nunu}, respectively.
\begin{table}[ht!]
\centering
\renewcommand{\arraystretch}{1} 
\begin{tabular}{|c|c|c|c|c|c|c|c|c|}
\hline
& &\multicolumn{2}{c}{$\Lambda_b\to \Lambda^*(\to p K^{-}) \mu^+\mu^- $ decay} & \multicolumn{2}{|c|}{$\Lambda_b\to \Lambda(\to p \pi) \mu^+\mu^-  $ decay}\\
\cline{2-6}
 Observables & $q^2$ bin & Central value  & $1\sigma$ range & Central value & $1\sigma$ range \\
\hline
  \hline
\multirow{2}{*} {BR} & {1.1 - 6.0} & 6.063 $\times 10^{-9}$& (4.660, 8.012) $\times 10^{-9}$& 0.775$\times 10^{-7}$ & (0.460, 1.164) $\times 10^{-7}$\\
                    & {14.2/15.0 - $\rm q^2_{max}$}  & 7.318 $\times 10^{-9}$&  (5.655, 9.100)$\times 10^{-9}$ & 3.723 $\times 10^{-7}$ & (3.105, 4.313 )$\times 10^{-7}$ \\
 \hline
 \multirow{2}{*} {$F_L$} & {1.1 - 6.0} & 0.781   &  (0.760, 0.800) &0.829 & (0.696, 0.907) \\
                    & {14.2/15.0 - $q^2_{max}$} & 0.430   & (0.424, 0.443) & 0.339 & (0.312, 0.375) \\
 \hline
  \multirow{2}{*} {$A_{FB}^L$} & {1.1 - 6.0} & -0.114  &  (-0.135, -0.089) &-0.028  & (-0.146, 0.051)  \\
                    & {14.2/15.0 - $q^2_{max}$}  &-0.236 &       (-0.274, -0.198)  & -0.299 &  (-0.330, -0.269)\\
 \hline
   \multirow{2}{*} {$\hat{K}_{1c}$} & {1.1 - 6.0} &    -0.152   &  (-0.180, -0.119) &  -0.019 &  (-0.097, 0.034) \\
                    & {14.2/15.0 - $q^2_{max}$} &   -0.313  & (-0.363, -0.262) &  -0.199 &   (-0.220, -0.180) \\
 \hline
    \multirow{2}{*} {$\hat{K}_{1cc}$} & {1.1 - 6.0} &  0.219   &  (0.200, 0.239) & 0.086 & (0.046, 0.152)\\
                    & {14.2/15.0 - $q^2_{max}$} &  0.565   &  (0.552, 0.573) & 0.331 &  (0.313, 0.344) \\
  \hline
    \multirow{2}{*} {$\hat{K}_{1ss}$} & {1.1 - 6.0} &  0.890   &  (0.880, 0.900) &  0.457 & (0.424, 0.477) \\
                    & {14.2/15.0 - $q^2_{max}$} &  0.713  & (0.719, 0.710) & 0.335 &  (0.328, 0.344) \\
  \hline
    \multirow{2}{*} {$\hat{K}_{2c}$} & {1.1 - 6.0} &  -0.038  &  (-0.045, -0.030) & 0.013 & (-0.012, 0.063) \\
                    & {14.2/15.0 - $q^2_{max}$} &  -0.080  &  (-0.093 ,-0.067) &  0.202 & (0.190, 0.210) \\
   \hline
    \multirow{2}{*} {$\hat{K}_{2cc}$} & {1.1 - 6.0} &  0.055  &  (0.050, 0.060) &  -0.054 & (-0.095, -0.029)\\
                    & {14.2/15.0 - $q^2_{max}$} & 0.145  &  (0.142, 0.146) &  -0.135 & (-0.149, -0.122) \\
     \hline
    \multirow{2}{*} {$\hat{K}_{2ss}$} & {1.1 - 6.0} &   0.223    &  (0.220, 0.225)  & -0.026  & (-0.048, -0.012) \\
                    & {14.2/15.0 - $q^2_{max}$} & 0.181      &  (0.180, 0.182) & -0.067 & (-0.074, -0.061)\\
     \hline
    \multirow{2}{*} {$\hat{K}_{4ss}$} & {1.1 - 6.0} &  0.000   & (-0.001, 0.001) &\bm{$\times$} &\bm{$\times$} \\
                    & {14.2/15.0 - $q^2_{max}$} &  -0.032   &  (-0.039, -0.027) &\bm{$\times$} & \bm{$\times$}\\
   \hline
   \multirow{2}{*} {$\hat{K}_{4sc}$} & {1.1 - 6.0} & \bm{$\times$}  & \bm{$\times$} &  0.003 &  (-0.062, 0.078) \\
                    & {14.2/15.0 - $q^2_{max}$} &\bm{$\times$} & \bm{$\times$} &  -0.043 & (-0.057, -0.030) \\
   \hline
   \multirow{2}{*} {$\hat{K}_{4s}$} & {1.1 - 6.0} &\bm{$\times$} &\bm{$\times$} &   0.031 & (-0.057, 0.110) \\
                    & {14.2/15.0 - $q^2_{max}$} & \bm{$\times$} & \bm{$\times$}& -0.116 & (-0.132, -0.100) \\
   \hline
    \multirow{2}{*} {$\hat{K}_{5s}$} & {1.1 - 6.0} & 0.014   & (0.011,0.018) & \bm{$\times$} &\bm{$\times$} \\
                    & {14.2/15.0 - $q^2_{max}$} & 0.046   &  (0.039, 0.055)  &\bm{$\times$} &\bm{$\times$} \\
   \hline
 \hline
    \multirow{2}{*} {$R_{\Lambda^{(*)}}$} & {1.1 - 6.0} & 0.996 &  (0.996, 0.996) & 0.995 & (0.989, 1.007)\\
                    & {14.2/15.0 - $q^2_{max}$} & 0.993 & (0.993,0.993)  & 1.007 &  (1.006, 1.007) \\
 \hline
    \multirow{2}{*} {$\mathcal{Q}_{F_L}$} & {1.1 - 6.0} & -0.000 &  (-0.000, -0.000) & -0.011 &  (-0.017, -0.007)\\
                    & {14.2/15.0 - $q^2_{max}$} & -0.000 &   (0.000, 0.000) & 0.001 &   (0.001,0.001)\\
 \hline
    \multirow{2}{*} {$\mathcal{Q}_{A_{FB}}$} & {1.1 - 6.0} & 0.000 & (0.000, 0.000) & 0.001 & (-0.000,0.003)  \\
                    & {14.2/15.0 - $q^2_{max}$} & 0.001 &  (0.001, 0.001) & -0.000 &  (-0.000, -0.000) \\
 \hline 
\end{tabular}
\caption{SM predictions of branching ratio~(BR), longitudinal polarization fraction~$F_L$, lepton forward-backward asymmetry~$A_{FB}$, 
angular coefficients $\hat{K}_i'$s and ratio of branching ratio~$R_{\Lambda^{(*)}}$ for the $\Lambda_b\to\Lambda^*(\to pK^-)\ell^+\ell^-$ and 
$\Lambda_b\to\Lambda(\to p\pi)\ell^+\ell^-$ decay channels.}
\label{sm_rslt_lst}
\end{table}

\begin{table}[ht!]
\centering
\setlength{\tabcolsep}{10pt} 
\renewcommand{\arraystretch}{2.0} 
\begin{tabular}{|c||c|c||c|c|}
\hline
&\multicolumn{2}{c}{$\Lambda_b\to \Lambda^*(\to p K^{-}) \nu\bar{\nu} $ decay} & \multicolumn{2}{|c|}{$\Lambda_b\to \Lambda(\to p \pi) \nu\bar{\nu} $ decay}\\
\hline
 Observables & Central value  &  $1\sigma$ range & Central value  &  $1\sigma$ range\\
  \hline \hline 
 {BR$\times 10^{-6}$}     &   1.414  & (1.148, 1.743)  &    1.795  &   (1.406, 2.202)\\
 \hline
 {$F_L$}                  &   0.522  & (0.503, 0.547)  & 0.472   &  (0.395, 0.564)\\
 \hline
$K_{1c}$                  & -0.421   & (-0.440, -0.391) & -0.207  & (-0.165, -0.241) \\
 \hline 
$K_{1cc}$                 & 0.477    &  (0.452, 0.497) & 0.264  &  (0.218, 0.302) \\
 \hline
$K_{1ss}$                 &  0.760   &  (0.751, 0.773) & 0.368  &  (0.349, 0.391) \\
 \hline
$K_{2c}$                  &  -0.106  &  (-0.110, -0.098) &   0.170 &  (0.140, 0.194)\\
 \hline 
$K_{2cc}$                 &  0.120   &  (0.114, 0.125) & -0.133  &   (-0.155, -0.106 ) \\
 \hline  
$K_{2ss}$                 & 0.190    &  (0.188, 0.193)  &  -0.066 &  (-0.077, -0.053) \\
 \hline
 \hline
 $K_{4sc}$               &\bm{$\times$}&\bm{$\times$}&          -0.032 &   (-0.060, 0.000)  \\
\hline
$K_{4s}$                & \bm{$\times$} &\bm{$\times$} & -0.061  &  (-0.099, -0.025) \\
\hline 
$K_{4ss}$                 &  -0.008  & (-0.011, -0.006) & \bm{$\times$} & \bm{$\times$} \\
\hline
$K_{5s}$                  &   0.022  &   (0.019, 0.026) &\bm{$\times$} & \bm{$\times$} \\
\hline
\end{tabular}
\caption{ SM prediction of $\Lambda_b\to \Lambda^* (\to p K^{-})\nu\bar{\nu}$ and $\Lambda_b\to\Lambda(\to p\pi)\nu\bar{\nu}$ decay 
observables.}
\label{sm_rslt_lst_nunu}
\end{table}

Our observations are as follows.
\begin{itemize}
 \item  The branching ratio of $\Lb\to \Lst(\to pK^-)\mu^+\mu^-$ mode is found to be of $\mathcal O(10^{-9})$, while the branching ratio of 
$\Lambda_b\to\Lambda(\to p\pi)\mu^+\mu^-$ decay mode is observed to be of $\mathcal O(10^{-7})$.  
 \item The values of $F_L$, $A_{FB}^l$, $\hat{K}_{1c}$, $\hat{K}_{2c}$, $\hat{K}_{2ss}$, and $\hat{K}_{4ss}$ are observed to be lower at high 
$q^2$ bin compared to the values obtained in the low $q^2$ bin.
 \item In the case of the $\Lambda_b\to\Lambda^*(\to pK^-)\mu^+\mu^-$ decay mode, values of  $\hat{K}_{3s}$ and $\hat{K}_{4ss}$ are
zero in the low $q^2$ bin, whereas they are non-zero in the high $q^2$ bin.
\item We found the ratios $K_{1c}/K_{2c}$, $K_{1cc}/K_{2cc}$ and $K_{1ss}/K_{2ss}$ to be equal to $4$.
 \item  As expected, value of $R_{\Lambda^{(*)}}$ is very close to unity. 
\item The branching fraction of both $\Lb\to \Lst(\to p{K}^-)\nu\bar{\nu}$ and $\Lambda_b\to\Lambda(\to p\pi)\nu\bar{\nu}$ decay channels are 
found to be of $\mathcal O(10^{-6})$.
\item It is observed that the uncertainties in the di-neutrino channels are less than the uncertainty in di-lepton decay channels.
 \end{itemize}
For completeness we also report the branching ratio for the $\Lambda_b\to\Lambda(\to p\pi)\tau^+\tau^-$ mode to be 
$(1.9\pm0.43) \times 10^{-7}$ in $q^2 \in[15.0- q^2_{max}]$ which is quite similar to the value reported in ref~.\cite{Das:2018iap}. A slight
difference is observed due to the different choice of input parameters.

\subsection{Global fit}

\label{old_one}
Our primary objective in this work is to use a model-independent SMEFT formalism to investigate the effects of $b \to s\,l^+\,l^-$ anomalies 
on several baryonic $b \to s\,l^+\,l^-$ and $b\,\to\,s\,\nu\,\bar{\nu}$ transition decays.
The SMEFT coefficients for left chiral currents, namely $\widetilde{c}_{ql}^{(1)}$, $\widetilde{c}_{ql}^{(3)}$, and $\widetilde{c}_{Z}$ 
contribute to WCs $C_{9,10}$ in $b \to s\,l^+\,l^-$ and to $C_L$ in $b \to s\nu \bar{\nu}$ transitions decays. Similarly, the SMEFT 
coefficients for right chiral currents such as $\widetilde{c}_{dl}$ and $\widetilde{c}_{Z}^{\prime}$ are connected to $C_{9,10}^\prime$ in 
$b \to s\,l^+\,l^-$ and $C_R$ in $b\to s\nu \bar{\nu}$ transition decays.
We construct several $1D$ and $2D$ NP scenarios. For $1D$ NP scenario, we consider NP contribution from a single NP operator,
whereas, for $2D$ NP scenario, we consider NP contribution from two different NP operators simultaneously. We use a naive $\chi^2$ analysis
and determine the scenario that best explains the anomalies observed in $b \to s\,l^+\,l^-$ transition decays. To obtain the best-fit values 
of these NP Wilson coefficients, we use all the available $b \to s\,l^+\,l^-$ experimental data. We define our $\chi^2$ as follows
\begin{equation}
\chi^2= \sum_i  \frac{\Big ({\cal O}_i^{\rm th} -{\cal O}_i^{\rm exp} \Big )^2}{(\Delta {\cal O}_i^{\rm exp})^2+(\Delta {\cal O}_i^{\rm th})^2
}\,,
\end{equation}
where ${\cal O}_i^{\rm th}$ and ${\cal O}_i^{\rm exp}$ denote the theoretical and measured central values of each observables, respectively. 
The uncertainties associated with theory and experimental values are represented by $\Delta {\cal O}_i^{\rm th}$ and 
$\Delta {\cal O}_i^{\rm exp}$. 
In our $\chi^2$ analysis, we include total eight measurements, namely ($R_K$, ${R_{K^*}}_{[q^2=1.1 - 6]}$ (BELLE), ${R_{K^*}}_{[q^2=1.1 - 6]}$
 (BABAR), ${P_5^{\prime}}_{[q^2=4 - 6]}$, ${P_5^{\prime}}_{[q^2=4.3 - 6]}$, ${P_5^{\prime}}_{[q^2=4 - 8]}$, 
$\mathcal{B}(B_s, \to, \phi, \mu^+,\mu^-)$, 
and $\mathcal{B}(B_s \to \mu^+\mu^-)$). The best fit values and the corresponding allowed ranges of all the SMEFT coefficients for
various $1D$ and $2D$ scenarios are reported in Table.~\ref{tab_fits_olddata}. We also report the $\chi^2_{\rm min}$/d.o.f and the 
Pull$_{\rm SM}= \sqrt{\chi^2_{\rm SM}-\chi^2_{\rm NP}}$ for each scenario in Table.~\ref{tab_fits_olddata}.
We consider eight measured parameters in our $\chi^2$ analysis. Hence, the number of degrees of freedom~(d.o.f)
will be $8-1=7$ for each $1D$ NP scenario and $8-2 = 6$ for each $2D$ NP scenario. We first determine the $\chi^2_{\rm min}$/d.o.f in the SM
to be $5.578$, which determines the degree of disagreement between the SM prediction and the current experimental data.
In each case, the $\chi^2_{\rm min}$ value represents the best-fit value. We impose $\chi^2 \le 12.592$ constraint to obtain the allowed
range of each $1D$ NP coefficient at $95\%$ percent confidence level~(CL). Similarly, the allowed range for each $2D$ NP coefficient at the 
$95\%$ CL is obtained by imposing $\chi^2 \le 11.070$ constraint.

\begin{table}[htbp]
\centering
\setlength{\tabcolsep}{6pt} 
\renewcommand{\arraystretch}{1} 
\begin{tabular}{|c||c|c|c|}
\hline
SMEFT couplings                                      & Best fit          & $\chi^2_{\rm min}$/d.o.f          &  Pull$_{\rm SM}$ \\
\hline 
SM                      &   \bm{\textemdash}     &     5.578        &      \bm{\textemdash}   \\

\hline
{$\widetilde{c}_{ql}^{(1),(3)}$}     &   -0.495        &    2.953     &    1.620   \\
              $1\sigma \to$          &  [-8.683, 0.335]     &      &        \\
\hline
{$\widetilde{c}_{Z}$}                &   0.862   &       3.374     &   1.484      \\
                               $1\sigma \to$   & [-0.529, 9.599] &        &          \\
\hline
{$\widetilde{c}_{Z}^{\prime}$}   &   -0.114       &     6.728         &     \bm{\textemdash}    \\
                               $1\sigma \to$   &  [-1.106, 1.027]      &            &          \\
\hline
{$\widetilde{c}_{dl}$ }          &   -0.114      &     6.954       &   \bm{\textemdash}      \\
                                $1\sigma \to$    & [-0.696, 0.693]        &            &          \\
\hline
{$(\widetilde{c}_{ql}^{(1)},\widetilde{c}_{ql}^{(3)})$ }   &   (-9.444, 8.797) & 3.391     &   1.478 \\
                                       $1\sigma \to$       &  ([-9.999, 9.969], [-9.998, 9.937])       &     &   \\                                                
\hline
{$(\widetilde{c}_{ql}^{(1),(3)},\widetilde{c}_{Z})$ }      &  (-1.732, -1.608) &       3.211    &  1.539\\
            $1\sigma \to$   &  ([-8.951, 2.619], [-5.293, 8.713])         &            &          \\   
 
\hline
{$(\widetilde{c}_{ql}^{(1),(3)},\widetilde{c}_{dl})$ }          & (-0.660, 0.211)   &        4.324   &  1.120       \\
          $1\sigma \to$      & ([-8.457, 0.175],[-2.333, 1.128])          &            &          \\  
                                                                                
\hline
{$(\widetilde{c}_{ql}^{(1),(3)},\widetilde{c}_{Z}')$ }  &   (-3.774, -4.827)  &  1.402     &  2.044  \\
                              $1\sigma \to$      &   ([-8.431, 0.175], [-6.038, 5.537])        &            &          \\  
\hline
{$(\widetilde{c}_{Z},\widetilde{c}_{dl})$ } &   (0.969, 0.211)     &        4.679    &    0.948    \\
                                     $1\sigma \to$     &   ([-0.167, 4.647], [-0.749, 1.837])        &            &          \\  
\hline
{$(\widetilde{c}_{Z},\widetilde{c}_{Z}^{\prime})$ } &    (4.492, -4.057)     &  1.863          &   1.927   \\
                               $1\sigma \to$      &  ([-0.195, 6.346], [-5.175, 4.725])         &            &          \\  
\hline                                                                                   
{$(\widetilde{c}_{dl},\widetilde{c}_{Z}^{\prime})$ } &    (-0.105, -0.164)       &     8.395       &   \bm{\textemdash}       \\
                        $1\sigma \to$    &  ([-2.616, 1.341], [-1.634, 0.881])          &            &     \\ 
  \hline
  \hline
{$(\widetilde{c}_{ql}^{(1)}+\widetilde{c}_{ql}^{(3)}$}    &     -0.495      &       2.953     &   1.620      \\ 
                                     $1\sigma \to$        &     [0.343, 1.157]           &            &          \\ 
\hline 
{$(\widetilde{c}_{ql}^{(1)}+\widetilde{c}_{ql}^{(3)},\widetilde{c}_{Z})$}             &   (-1.119, -0.804)   &   3.383   & 
 1.482\\ 
                               $1\sigma \to$    &   ([-8.804, 2.655], [-5.422, 8.742])    &            &          \\ 
\hline 
{$(\widetilde{c}_{ql}^{(1)}+\widetilde{c}_{ql}^{(3)},\widetilde{c}_{dl})$}   &   ( -0.645, -0.010)       &      3.671    
& 1.381        \\ 
               $1\sigma \to$       &  ([-8.453, 0.157], [-2.323, 1.084])         &            &          \\
\hline 
{$(\widetilde{c}_{ql}^{(1)}+\widetilde{c}_{ql}^{(3)},\widetilde{c}_{Z}^{\prime})$}   &   (-3.776, -4.938)     & 
1.417  &    2.040     \\ 
                      $1\sigma \to$            & ([-8.447, 0.157], [-6.078, 5.545] )        &            &          \\
\hline 
\end{tabular}
\caption{Best fit values and the allowed ranges of SMEFT coefficients at $95\%$ CL in several $1D$ and $2D$ scenarios.}
\label{tab_fits_olddata}
\end{table}
It is evident from Table~\ref{tab_fits_olddata} that not all the SMEFT coefficients can explain the observed deviations in
$b \to s\,l^+\,l^-$ data. In fact, NP scenarios represented by $\widetilde{c}_{dl}$, $\widetilde{c}_{Z}^{\prime}$ and
$(\widetilde{c}_{dl}$, $\widetilde{c}_{Z}^{\prime})$ WC's are ruled out because the $\chi^2_{min}$ values
obtained for these scenarios are higher than the $\chi^2$ value obtained in SM.
Hence, we will not discuss them any further. Nevertheless, there are a few $2D$ scenarios, namely
$(\widetilde{c}_{ql}^{(3)},\widetilde{c}_{Z}^{\prime})$, $(\widetilde{c}_{Z},\widetilde{c}_{Z}^{\prime})$,
and $(\widetilde{c}_{ql}^{(1)}+\widetilde{c}_{ql}^{(3)},
\widetilde{c}_{Z}^{\prime})$, for which the Pull$_{\rm SM}$ is
considerably larger
than the rest of the NP scenarios. Furthermore, these scenarios exhibit better compatibility with $R_K$, $R_{K^*}$,
$P_5^{\prime}$, $\mathcal{B}(B_s \to \phi \mu^+\mu^-)$ and $\mathcal{B}(B_s \to \mu^+\mu^-)$ data.
In Table.~\ref{tab_gen_smeft}, we present the central values and the corresponding $1\sigma$ uncertainties associated
with each observable pertaining to $B \to , K^{(*)}\,\mu^{+}\mu^{-}$ decays in the SM and in case of several NP scenarios.
The experimental values till $2022$ December for $R_K$, ${R_{K^*}}_{[q^2=1.1 - 6]}$, ${P_5^{\prime}}_{[q^2=4 - 6]}$,
${P_5^{\prime}}_{[q^2=4.3 - 6]}$, ${P_5^{\prime}}_{[q^2=4 - 8]}$, $\mathcal{B}(B_s \to \phi \mu^+\mu^-)$, and
$\mathcal{B}(B_s \to \mu^+\mu^-)$ are also listed in the first row of Table.~\ref{tab_gen_smeft}.

 \begin{table}[htbp]
\setlength{\tabcolsep}{5pt} 
\renewcommand{\arraystretch}{1} 
\resizebox{\columnwidth}{!}{
\begin{tabular}{|c||c|c|c|c|c|c|c|}
\hline
SMEFT couplings     & $R_K$  & $R_{K^*}$ & $P'_5$ [4.0,6.0] & $P'_5$ [4.3, 6.0] & $P'_5$ [4.0, 8.0]
& $\mathcal{B}(B_s\to \phi \mu \mu)\times 10^{-7}$ & $\mathcal{B}(B_s \to \mu \mu)\times 10^{-9}$\\
\hline 
Expt. Value      &      0.846 $\pm$ 0.060       &      $0.685 \pm 0.150$            &   -0.21$\pm$ 0.15     &      -0.96$\pm$ 0.272     &  -0.267$\pm$0.279   &       1.44$\pm$ 0.21    &  3.09$\pm$ 0.484  \\ 
\hline 
$\widetilde{c}_{ql}^{(3)}$  & 0.792  &   0.791  &  -0.693  &  -0.702  &  -0.719  &   2.314  &   3.108\\
$1\sigma$                       &  (0.343, 1.157)  & (0.359, 1.352)   & (-0.985, -0.500)   & (-0.986, -0.524)  &  (-0.991, -0.591)  &  (1.089, 4.127)  & (1.179, 4.549) \\

\hline
  $\widetilde{c}_{Z}$ & 0.810  &   0.784   & -0.768   & -0.775&    -0.778  &   2.227  &   2.548\\
 $1\sigma$  &(0.429, 1.146) & (0.348, 1.151) & (-1.062, 0.761)  & (-1.064, 0.770) &  (-1.040, 0.793) & (1.065, 4.094) &(0.146, 4.869) \\

  \hline 
  {$(\widetilde{c}_{ql}^{(1)},\widetilde{c}_{ql}^{(3)})$} & 0.737  &   0.731  &  -0.672 &   -0.682  &  -0.705  &   2.190 &    2.883\\
        $1\sigma$                              & (0.412, 1.094) & (0.421, 1.185) & (-0.975, -0.535) & (-0.976, -0.552)& (-0.980, -0.603)& (1.269, 3.827) & (1.575, 4.290)\\
  \hline 
  {$(\widetilde{c}_{ql}^{(1),(3)},\widetilde{c}_{Z})$ }  & 0.694 &    0.751  &  -0.491  &  -0.508  &  -0.569 &    2.160    & 3.622\\
               $1\sigma$       & (0.363, 1.102)& (0.427, 1.218) & (-1.041, 0.530) & (-1.042, 0.544) & (-1.020, 0.598) & (1.275, 3.721) &(0.000, 6.022)\\

   \hline  
   {$(\widetilde{c}_{ql}^{(1),(3)},\widetilde{c}_{Z}')$ } & 0.825  &   0.594 &   -0.329 &   -0.314  &  -0.255  &   1.756  &   2.838  \\
          $1\sigma$       & (0.353, 1.330) & (0.223, 1.151) & (-1.170, 0.518) & (-1.178, 0.507)  & (-1.179, 0.464) & (0.643, 3.799) &(0.000, 5.717)\\
        
   \hline 
 {$(\widetilde{c}_{Z},\widetilde{c}_{Z}^{\prime})$ } & 0.900   &  0.709  &  -0.311 &   -0.295  &  -0.227 &    2.073 &    2.717\\
             $1\sigma$                & (0.379, 1.300)& (0.351, 1.131) & (-1.187, 0.592) &(-1.186, 0.576) & (-1.149, 0.502) & (1.054, 3.910)
                                                    & (0.023,5.466 ) \\
\hline 
 $(\widetilde{c}_{ql}^{(1)}+\widetilde{c}_{ql}^{(3)},\widetilde{c}_{Z})$ & 0.733   &  0.754  &  -0.589   & -0.601  &  -0.637  &   2.327    & 3.494 \\
                               $1\sigma$             & (0.363, 1.109) & (0.422, 1.161) &(-1.012, 0.532) & (-1.013, 0.546) &(-1.000, 0.605) & (1.255, 3.730) &(0.000, 5.993) \\
                    
 \hline 
  $(\widetilde{c}_{ql}^{(1)}+\widetilde{c}_{ql}^{(3)},\widetilde{c}_{dl})$  & 0.734  &   0.737  &  -0.661   & -0.672&    -0.701  &   2.322 &    2.851 \\
                               $1\sigma$             & (0.387, 1.205) & (0.271, 1.135)  & (-1.141, -0.161) & (-1.146, -0.161) & (-1.146, -0.159)  &  (-1.166, -0.013)  & (0.804, 3.679) \\
 \hline 
 $(\widetilde{c}_{ql}^{(1)}+\widetilde{c}_{ql}^{(3)},\widetilde{c}_{Z}^{\prime})$ & 0.855 &    0.611 &   -0.280  &  -0.266 &   -0.211    & 1.785 &    3.006\\
                                         $1\sigma$                                           & (0.360, 1.317) & (0.215, 1.163) & (-1.173, 0.522) &(-1.179, 0.511) 
                                                                                &(-1.179, 0.468) & (0.664, 3.757) & (0.000, 5.868)\\
\hline 
\end{tabular}}
\caption{Best fit values and the corresponding allowed ranges of $R_K$, $R_{K^*}$, $P'_5$ [4.0,6.0],  $P'_5$ [4.3, 6.0], $P'_5$ [4.0, 8.0]
, $\mathcal{B}(B_s\to \phi \mu^+ \mu^-)$ and $\mathcal{B}(B_s \to \mu^+\mu^-)$ with each NP scenarios of Table.~\ref{tab_fits_olddata}.}
\label{tab_gen_smeft}
\end{table}
We now move to analyse the goodness of our fit results with the measured values of $\mathcal B(B \to \, K^{(*)} \, \nu\,\bar{\nu})$.
We report, in Table.~\ref{tab_b2knn_fit}, the best fit values and the corresponding allowed ranges of  
$\mathcal{B}(B \to \, K^{(*)} \, \nu\,\bar{\nu})$, $F_{L}^{K^*}$ and also the ratios $\mathcal{R_K}$, $\mathcal{R_{K^*}}$
and $\mathcal{R_{F_L}^{K^*}}$ obtained with the best fit values and the allowed ranges of each NP couplings at $95\%$ CL of 
Table.~\ref{tab_fits_olddata}. We also report the SM central values and the corresponding $1\sigma$ uncertainties associated with each 
observable in Table.~\ref{tab_b2knn_fit}. 
In the SM, the branching fractions of $B \to \, K^{(*)} \, \nu\,\bar{\nu}$ decays are of $\mathcal{O}(10^{-6})$.
The ratios $\mathcal{R_K}$, $\mathcal{R_{K^*}}$ and $\mathcal{R_{F_L}^{K^*}}$ are equal to unity in the SM.
Hence any deviation from unity in these parameters could be a clear signal of beyond the SM physics.
Moreover, there exists a few experiments that provide the upper bound on the branching ratio of $B \to K^{(*)}\,\nu\,\bar{\nu}$ to be
$\mathcal{B}({B \to K\,\nu\,\bar{\nu}}) < 11 \times 10^{-6}$ and $\mathcal{B}({B \to K^*\,\nu\,\bar{\nu}}) < 27 \times 10^{-6}$, respectively.
Ignoring any theoretical uncertainty, we estimate the upper bound on $\mathcal{R_K^{(*)}}$ to be $\mathcal{R_K} < 2.75$ and
$\mathcal{R_{K^*}} < 2.89$, respectively.

We observe that the range of $B \to K^{(*)}\,\nu\,\bar{\nu}$ and $\mathcal{R_K^{(*)}}$ obtained with the allowed range of each NP couplings
are compatible with the experimental upper bound of  $B \to K^{(*)}\,\nu\,\bar{\nu}$ and $\mathcal{R_K^{(*)}}$. However, the best fit values
of $B \to K^{(*)}\,\nu\,\bar{\nu}$ and $\mathcal{R_K^{(*)}}$ obtained with the best fit values of 
$(\widetilde{c}_{ql}^{(1)},\widetilde{c}_{ql}^{(3)})$, $(\widetilde{c}_{ql}^{(1)},\widetilde{c}_Z^{\prime})$ and 
$(\widetilde{c}_{ql}^{(1)} + \widetilde{c}_{ql}^{(3)},\widetilde{c}_Z^{\prime})$ SMEFT coefficients are larger than the experimental
upper bound. Hence a simulateneous explanation of $b \to s\,l^+\,l^-$ and $b \to s\nu\bar{\nu}$ is not possible with these NP scenarios.
Moreover, the values of $B \to K^{(*)}\,\nu\,\bar{\nu}$ and $\mathcal{R_K^{(*)}}$ obtained with 
$(\widetilde{c}_{ql}^{(1)},\widetilde{c}_Z)$ SMEFT coefficients are quite large. More precise measurement on 
$B \to K^{(*)}\nu\bar{\nu}$ branching fraction in future can exclude this NP scenario. 

Again it can be seen from Table.~\ref{tab_b2knn_fit} that $\mathcal{R_{F_L}^{K^*}}$ remains SM like for all the scenarios with left handed currents. 
However, with the inclusion of right handed currents, its value seem to differ from unity. Hence a deviation from unity in 
$\mathcal{R_{F_L}^{K^*}}$ would be clear signal of NP through right handed currents. It should be emphasized that the value of 
$\mathcal{R_{F_L}^{K^*}}$ obtained with $(\widetilde{c}_{ql}^{(1)},\widetilde{c}_Z^{\prime})$, 
$(\widetilde{c}_{ql}^{(3)},\widetilde{c}_Z^{\prime})$, $(\widetilde{c}_Z,\widetilde{c}_Z^{\prime})$, and 
$(\widetilde{c}_{ql}^{(1)} + \widetilde{c}_{ql}^{(3)},\widetilde{c}_Z^{\prime})$ SMEFT couplings deviates significantly from the SM 
prediction. 

\begin{table}[htbp]
\centering
\setlength{\tabcolsep}{8pt} 
\renewcommand{\arraystretch}{1.5} 
\resizebox{\columnwidth}{!}{
\begin{tabular}{|c||c|c|c|c|c|c|}
\hline
SMEFT couplings & {$\mathcal{B}({B \to\, K \, \nu\,\bar{\nu}})\times 10^{-6}$} & {$\mathcal{R_K}$} & {$\mathcal{B}({B \to\, K^* \, \nu\,\bar{\nu}})\times 10^{-6}$} & {$\mathcal{R_{K^*}}$} & {$F_L ({B \to\, K^* \, \nu\,\bar{\nu}})$} & {$\mathcal{R_{F_L}^{K^*}}$} \\
\hline
\hline
SM          & {$4.006 \pm 0.261$} & {1.000} & {$9.331 \pm 0.744$} & {1.000} & {$0.493 \pm 0.038$} & {1.000}  \\ 
\hline
{$\widetilde{c}_{ql}^{(1)}$} &     4.490 &    1.162  &  10.690  &   1.162  &   0.427  &   1.000    \\ 
$1\sigma \to $ & (3.105, 25.688)  & (0.897, 5.602)   &  (6.974, 62.949)   &  (0.897, 5.602)  & (0.372, 0.617)  & (1.000, 1.000)   \\
\hline
{$\widetilde{c}_{ql}^{(3)}$} & 3.285  &   0.850   &  7.821  &   0.850   &  0.427  &   1.000\\ 
$1\sigma \to $ & (0.071, 5.079)  & (0.020, 1.108)    &  (0.181, 12.116)   & (0.020, 1.108)   & (0.372, 0.617)  & (1.000, 1.000)   \\
\hline
{$\widetilde{c}_{Z}$}      & 3.283  &   0.747  &   7.095  &   0.747  &   0.465  &   1.000   \\ 
$1\sigma \to $ & (0.364, 5.135)  & (0.090, 1.174)   & (0.796, 12.854)    & (0.090, 1.174)   & (0.374, 0.631)  &  (1.000, 1.000) \\
\hline
 $(\widetilde{c}_{ql}^{(1)},\widetilde{c}_{ql}^{(3)})$ & 61.985 &   14.989 &  160.627  &  14.989   &  0.488 & 1.000\\ 
 $1\sigma \to $ & (0.000, 76.129)  & (0.000, 17.069)   &  (0.000, 179.628)   & (0.000, 17.069)   & (0.366, 0.614) 
 &  (1.000, 1.000)  \\
\hline
$(\widetilde{c}_{ql}^{(1)},\widetilde{c}_{Z})$ & 9.349  &   2.328  &  19.468  &   2.328  &   0.458  &   1.000\\ 
$1\sigma \to $ & (0.000, 28.557) & (0.000, 7.123)   & (0.000, 72.582)    & (0.000, 7.123)   & (0.368, 0.617)  &   (1.000, 1.000) \\
\hline
$(\widetilde{c}_{ql}^{(3)},\widetilde{c}_{Z})$ & 3.860 &    0.961   &  8.038   &  0.961   &  0.458 &    1.000
\\ 
$1\sigma \to $ &  (0.053, 6.019) & (0.015, 1.387)   &  (0.143, 14.583)   &  (0.015, 1.387)  & (0.368, 0.617)  & (1.000, 1.000)    \\
\hline
$(\widetilde{c}_{ql}^{(1)},\widetilde{c}_{dl})$ & 4.784 &    1.146  &  12.868    & 1.271   &  0.471   &  1.017 \\ 
$1\sigma \to $ &  (2.991, 26.517)  &  (0.868, 6.037)  & (7.790, 58.665)    &  (0.940, 5.291)  & (0.348, 0.622)  
&   (0.822, 1.088) \\
\hline
$(\widetilde{c}_{ql}^{(3)},\widetilde{c}_{dl})$ & 3.108  &   0.745  &   8.565  &   0.846   &  0.472  &   1.021
 \\ 
$1\sigma \to $ & (0.000, 4.735)  & (0.000, 1.088)   &  (0.373, 10.768)   & (0.045, 1.084)   & (0.362, 0.685)  & 
(0.893, 1.248)\\
\hline
$(\widetilde{c}_{ql}^{(1)},\widetilde{c}_{Z}^{\prime})$ & 22.421   &  5.541  &  12.148  &   1.527  &   0.236  &   0.456  \\ 
$1\sigma \to $ & (2.991, 30.806)  & (0.812, 6.835)   & (7.484, 79.139)    &  (0.863, 6.891)  & (0.113, 0.685)  & (0.262, 1.224)   \\
\hline
$(\widetilde{c}_{ql}^{(3)},\widetilde{c}_{Z}^{\prime})$ & 5.499    & 1.359 &    2.681  &   0.337&     0.193   &  0.372  \\ 
$1\sigma \to $ & (0.140, 8.659)  & (0.037, 1.951)   &  (0.140, 10.852)   & (0.016, 1.150)   & (0.000, 0.665)  &  (0.000, 1.170)  \\
\hline
$(\widetilde{c}_{Z},\widetilde{c}_{dl})$ &2.361   &  0.663  &   6.629 &    0.759  &   0.469  &   1.022 \\ 
$1\sigma \to $ & (0.001, 4.700)  & (0.000, 1.149)   & (1.481, 11.080)    &  (0.179, 1.076)  & (0.364, 0.705)  & (0.862, 1.266)    \\
\hline
$(\widetilde{c}_{Z},\widetilde{c}_{Z}^{\prime})$ & 3.191  &   0.868  &   2.312  &   0.238  &   0.258 &    0.502\\ 
$1\sigma \to $ &  (0.000, 6.147) & (0.000, 1.375)    &  (0.398, 11.688)   &  (0.042, 1.128)  & (0.000, 0.706)  &  (0.000, 1.269)  \\
\hline
$(\widetilde{c}_{ql}^{(1)}+\widetilde{c}_{ql}^{(3)},\widetilde{c}_{Z})$ &5.234   &  1.269  &  12.505 &    1.269  &   0.431   &  1.000  \\ 
$1\sigma \to $ & (0.000, 14.409)  &  (0.000, 3.436)  &  (0.000, 39.087)   & (0.000, 3.436)   & (0.371, 0.612)  &  (1.000, 1.000)  \\
\hline
$(\widetilde{c}_{ql}^{(1)}+\widetilde{c}_{ql}^{(3)},\widetilde{c}_{dl})$ &4.091  &   1.003   &  9.883    & 0.998  &   0.509  &   0.999   \\ 
$1\sigma \to $ & (2.415, 8.104)  &  (0.688, 1.865)  &   (5.307, 13.180)  & (0.633, 1.283)   & (0.294, 0.635)  & (0.617, 1.100)   \\
\hline
$(\widetilde{c}_{ql}^{(1)}+\widetilde{c}_{ql}^{(3)},\widetilde{c}_{Z}^{\prime})$ & 12.906   &  3.159 &    5.752 &    0.586    & 0.042    & 0.085   \\ 
 $1\sigma$ &  (0.060, 17.140)  &  (0.016, 3.829)   &  (4.165, 31.904)  &  (0.416, 3.068)   &  (0.001, 0.708)    & 
 (0.003, 1.285)\\
\hline
\end{tabular}}
\caption{Best fit values and the corresponding allowed ranges of $\mathcal{B}(B \to \, K^{(*)} \, \nu\,\bar{\nu})$, $F_{L}^{K^*}$ and the
ratios $\mathcal{R_K}$, $\mathcal{R_{K^*}}$ and $\mathcal{R_{F_L}^{K^*}}$ in SM and in the presence of NP scenarios of 
Table.~\ref{tab_fits_olddata}.}
\label{tab_b2knn_fit}
\end{table}

\begin{figure}[htbp]
\centering
\includegraphics[width=8cm,height=5cm]{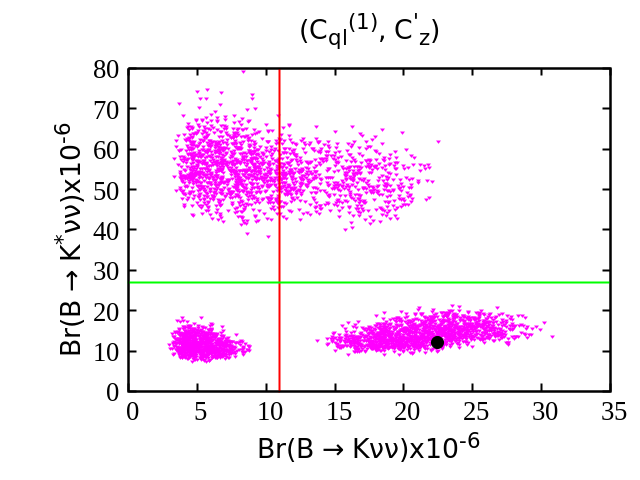}
\includegraphics[width=8cm,height=5cm]{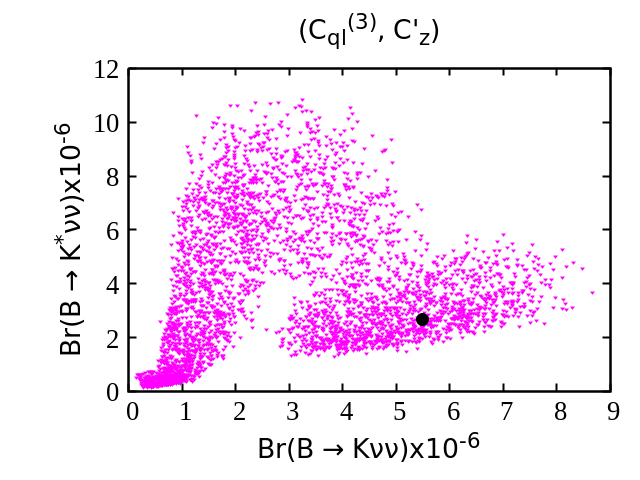}
\includegraphics[width=8cm,height=5cm]{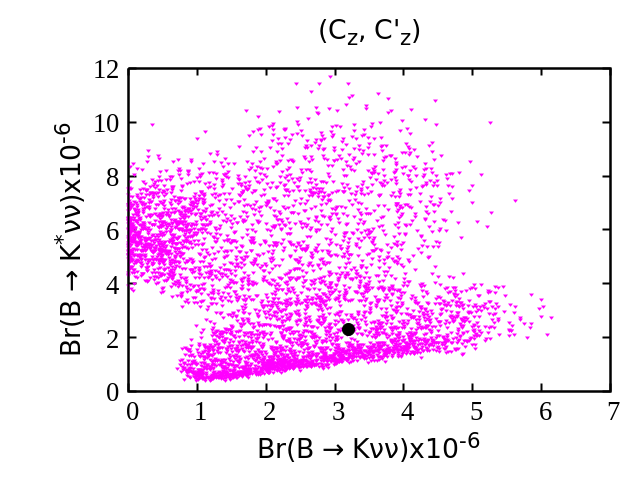}
\includegraphics[width=8cm,height=5cm]{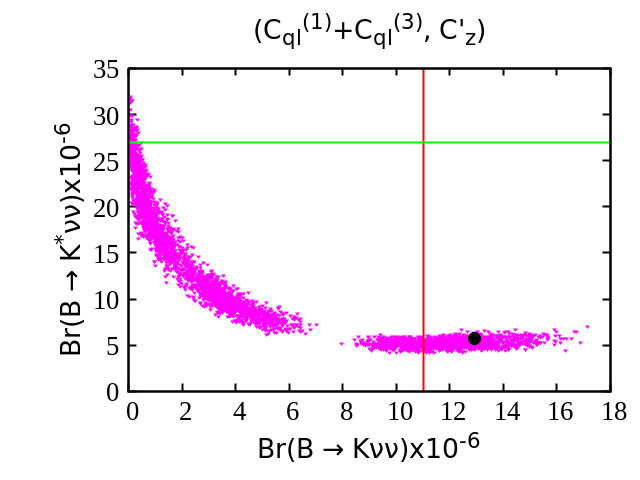}
\caption{Best fit values~(black dot) and the corresponding allowed ranges of $\mathcal{B}(B \to \, K \, \nu\,\bar{\nu})$ and 
$\mathcal{B}(B \to \, K^{*} \, \nu\,\bar{\nu})$ in case of $(\widetilde{c}_{ql}^{(1)},\widetilde{c}_Z^{\prime})$,
$(\widetilde{c}_{ql}^{(3)},\widetilde{c}_{Z}^{\prime})$, $(\widetilde{c}_{Z},\widetilde{c}_{Z}^{\prime})$,
and $(\widetilde{c}_{ql}^{(1)}+\widetilde{c}_{ql}^{(3)}, \widetilde{c}_{Z}^{\prime})$ NP scenarios. 
The red and green line represents the experimental upper bound of
$\mathcal{B}(B \to \, K \, \nu\,\bar{\nu})< 11 \times 10^{-6}$ and $\mathcal{B}(B \to \, K^{*} \, \nu\,\bar{\nu}) < 27 \times 10^{-6}$, 
respectively.}
\label{scatter_nunu}
\end{figure}

In Fig.~\ref{scatter_nunu}, we show the allowed ranges of  $\mathcal{B}(B \to \, K^{(*)} \, \nu\,\bar{\nu})$ with few selected NP scenarios
such as $(\widetilde{c}_{ql}^{(1)},\widetilde{c}_Z^{\prime})$,
$(\widetilde{c}_{ql}^{(3)},\widetilde{c}_{Z}^{\prime})$, $(\widetilde{c}_{Z},\widetilde{c}_{Z}^{\prime})$,
and $(\widetilde{c}_{ql}^{(1)}+\widetilde{c}_{ql}^{(3)},
\widetilde{c}_{Z}^{\prime})$ that best explains the $b \to s\,l^+\,l^-$ data. Best fit values of 
$\mathcal{B}(B \to \, K^{(*)} \, \nu\,\bar{\nu})$ are shown with a black dot in Fig.~\ref{scatter_nunu}. The allowed range of each observable
is obtained by using the allowed ranges of the NP couplings. The red and green line represents the experimental upper bound of 
$\mathcal{B}(B \to \, K \, \nu\,\bar{\nu})$ and $\mathcal{B}(B \to \, K^{*} \, \nu\,\bar{\nu})$, respectively. It is evident that
the allowed ranges of $\mathcal{B}(B \to \, K \, \nu\,\bar{\nu})$ and $\mathcal{B}(B \to \, K^{*} \, \nu\,\bar{\nu})$ obtained with
$(\widetilde{c}_{ql}^{(3)},\widetilde{c}_{Z}^{\prime})$ and $(\widetilde{c}_{Z},\widetilde{c}_{Z}^{\prime})$ SMEFT scenarios are compatible
with the experimental upper bound. In case of $(\widetilde{c}_{ql}^{(1)},\widetilde{c}_Z^{\prime})$ and 
$(\widetilde{c}_{ql}^{(1)}+\widetilde{c}_{ql}^{(3)}, \widetilde{c}_{Z}^{\prime})$ NP scenarios, although the best fit value does not 
simultaneously satisfy the experimental upper bound, there still exist some NP parameter space that can, in principle, satisfy both the
constraint. It is also evident that, the best fit value of $\mathcal{B}(B \to \, K \, \nu\,\bar{\nu}) = 12.9 \times 10^{-6}$ obtained
with $(\widetilde{c}_{ql}^{(1)}+\widetilde{c}_{ql}^{(3)}, \widetilde{c}_{Z}^{\prime})$ NP coupling is very close to the experimental upper
bound of $11 \times 10^{-6}$.

\subsection{Effects of SMEFT coefficients in $\Lb\to \Lst(\to p{K}^-)\mu^+\mu^{-}$ and  $\Lambda_b\to\Lambda(\to p\pi)\mu^+\mu^{-} $ decay 
observables}
Our main objective is to investigate NP effects on $\Lb\to \Lst(\to p{K}^-)\mu^+\mu^{-}$ and  $\Lambda_b\to\Lambda(\to p\pi)\mu^+\mu^{-} $ 
decay observables in a model independent SMEFT framework. Based on our $\chi^2$ analysis and the constraint imposed by the 
experimental upper bound of $\mathcal{B}(B \to \, K \, \nu\,\bar{\nu})$ and $\mathcal{B}(B \to \, K^{*} \, \nu\,\bar{\nu})$,
we chose three NP scenarios namely, 
$(\widetilde{c}_{ql}^{(3)},\widetilde{c}_{Z}^{\prime})$, $(\widetilde{c}_{Z},\widetilde{c}_{Z}^{\prime})$,
and $(\widetilde{c}_{ql}^{(1)}+\widetilde{c}_{ql}^{(3)}, \widetilde{c}_{Z}^{\prime})$ that corresponds to larger Pull$_{\rm SM}$ value than 
the rest of the NP scenarios. The results are listed in
Table.~\ref{Lsmumu_np_1}, Table.~\ref{Lsmumu_np_2}, Table.~\ref{Lmumu_np_1} and Table.~\ref{Lmumu_np_2}, respectively.
Our observations are as follows

\begin{table}[htbp]
\centering
\setlength{\tabcolsep}{4pt} 
\renewcommand{\arraystretch}{1} 
\begin{tabular}{|c|c|c|c|c|c|c|c|c|}
\hline 
 \multicolumn{6}{|c|}{$\Lambda_b\to \Lambda^*(\to p K^{-}) \mu^{+} \mu^{-}$ decay ($\mu$ mode)}\\
\hline 
\hline 
\multicolumn{2}{|c|}{SMEFT Couplings} & {BR$\times 10^{-9}$} & {$F_L$} &{$A_{FB}^{\mu}$} &{$R_{\Lambda^{*}}$}\\
  \hline 
  \hline

    \multirow{6}{*}{ $(\widetilde{c}_{ql}^{(3)},\widetilde{c}_{Z}^{'})$}  & $1.1-6.0$& 4.785  &   0.706 &  0.056  &0.786   \\
                                                    &  & [0.294, 14.542] & [0.051, 0.804] & [-0.276, 0.316] & [0.014, 2.100]   \\
                                                  
  \cline{2-6}
                                                    &$14.2-q^2_{max}$ & 3.972   & 0.477  &  0.028   &0.539     \\
                                                    &  & [0.263, 22.544] & [0.325, 0.558] & [-0.351, 0.091] & [0.008, 4.414]  \\

  \hline 
    \multirow{4}{*}{ $(\widetilde{c}_{Z},\widetilde{c}_{Z}^{\prime})$} & $1.1-6.0$&   5.102 &  0.802 &  0.017&0.838  \\
                                                    &  & [1.453, 12.108] & [0.738, 0.910] & [-0.131, 0.087] & [0.085, 1.782]   \\
                                                 
  \cline{2-6}
                                                    &$14.2-q^2_{max}$ & 4.921 &  0.393 &  0.001  & 0.668 \\
                                                    & & [2.085, 19.113] & [0.344, 0.528] & [-0.352, 0.201]  & [0.075, 3.464]  \\
                                                    
    \hline
     \multirow{6}{*}{  $(\widetilde{c}_{ql}^{(1)}+\widetilde{c}_{ql}^{(3)},\widetilde{c}_{Z}^{\prime})$} & $1.1-6.0$&   4.990 &  0.707 &  0.055  & 0.820 \\
                                                   & & [0.293, 14.580] & [0.048, 0.804] & [-0.277, 0.314]   & [0.013, 2.124]  \\ 
                                                 
  \cline{2-6}
                                                    &$14.2-q^2_{max}$ & 4.165 & 0.476 &  0.029 & 0.565  \\
                                                    &  & [0.266, 22.458] & [0.559, 0.325] & [-0.351, 0.093] & [0.008,4.397] \\
                                               
\hline 
 \end{tabular}
 \caption{The branching ratio~(BR), longitudinal polarization fraction $F_L$, lepton forward backward asymmetry $A_{FB}^{\mu}$ and the
ratio of branching ratio $R_{\Lambda^{*}}$ for the $\Lambda_b\to \Lambda^*(\to p K^{-}) \mu^{+} \mu^{-}$ decay mode in case of few selected
$2D$ NP scenarios.}
\label{Lsmumu_np_1}
 \end{table}

\begin{table}[htbp]
\centering
\setlength{\tabcolsep}{4pt} 
\renewcommand{\arraystretch}{1.38} 
\scalebox{0.45}{}
\resizebox{\columnwidth}{!}{
\begin{tabular}{|c|c|c|c|c|c|c|c|c|c|c|}
\hline 
 \multicolumn{10}{|c|}{$\Lambda_b\to \Lambda^*(\to p K^{-}) \mu^{+} \mu^{-}$ decay ($\mu$ mode)}\\
\hline 
\hline 
SMEFT Couplings & {$q^2$ bin} &{$K_{1c}$} & {$K_{1cc}$} &{$K_{1ss}$}   &{$K_{2c}$} & {$K_{2cc}$} &{$K_{2ss}$}
&{$K_{4ss}$}&{$K_{5s}$}\\
  \hline

  \hline 
     \multirow{6}{*}{ $(\widetilde{c}_{ql}^{(3)},\widetilde{c}_{Z}^{'})$}  & $1.1-6.0$& 0.074  & 0.294 &  0.853&   0.019 &  0.074 &  0.213   
     & -0.000 & -0.007\\
                                                     &  & [-0.369, 0.422] & [0.196, 0.949] & 
                                                     [0.525, 0.902]
                                                     & [-0.092, 0.105] & [0.049, 0.237] & [0.131, 0.225] & [-0.001, 0.001] &
                                                     [-0.010, 0.023]\\

   \cline{2-10}
                                                     &$14.2-q^2_{max}$  &  0.518 &  0.736  & 0.009 &  0.134 &  0.182  & 0.008  & -0.046 &  0.539  \\
                                                     & & [-0.465, 0.123] & [0.430, 0.672] &  [0.661, 0.773]
                                                     & [-0.119, 0.030] & [0.116, 0.170] &
                                                     [0.160, 0.185]
                                                      & [-0.049, 0.036] & [-0.054, 0.067] \\

    \hline 
     \multirow{6}{*}{ $(\widetilde{c}_{Z},\widetilde{c}_{Z}^{\prime})$} & $1.1-6.0$& 0.023 &  0.198 &  0.901 &  0.006 &  0.050&   0.225  &  0.000 & -0.005\\
                                                     &  & [-0.174, 0.116] & [0.090, 0.262] & [0.869, 0.955]   & [-0.044, 0.029] & [0.023, 0.066] &
                                                     [0.215, 0.239]  &  [-0.001, 0.001] & [-0.013, 0.017]\\
                                                   
   \cline{2-10}
                                                     &$14.2-q^2_{max}$ &  0.001 & 0.601 & 0.694 & 0.000 &   0.155& 0.175 & -0.002& -0.016 \\
                                                     & & [-0.467, 0.266] & [0.459, 0.653] & [0.670, 0.758]  & [-0.119, 0.068] & [0.124, 0.166] 
                                                     & [0.169, 0.188]  & [-0.047, 0.007] & [-0.054, 0.065]\\

     \hline
      \multirow{6}{*}{  $(\widetilde{c}_{ql}^{(1)}+\widetilde{c}_{ql}^{(3)},\widetilde{c}_{Z}^{\prime})$} & $1.1-6.0$&0.073 &0.293 & 0.854 &0.018 &0.073 
      &0.213&                                                                              -0.000&-0.007 \\
                                                     &  & [-0.369, 0.419] & [0.196, 0.951] & [0.524, 0.902]
                                                     & [-0.092, 0.105] & [0.049, 0.238] & [0.131, 0.225]  & [-0.001, 0.001]& [-0.010, 0.023] \\
                                                   
   \cline{2-10}
                                                     &$14.2-q^2_{max}$ &   0.038 & 0.519&0.735   &0.010 &  0.134 & 0.182 & 0.030 & -0.046\\
                                                     & & [-0.465, 0.126] & [0.429, 0.671] & [0.661, 0.773]
                                                     & [-0.119, 0.030] & [0.116, 0.170] & [0.160, 0.185]  & [-0.049, 0.036] & [-0.049, 0.067]\\

\hline 
 \end{tabular}}
 \caption{Angular observables $K_i$ for the $\Lambda_b\to \Lambda^*(\to p K^{-}) \mu^{+} \mu^{-}$ decay mode 
in case of few selected $2D$ NP scenarios.}
\label{Lsmumu_np_2}
 \end{table}


\begin{table}[htbp]
\centering
\setlength{\tabcolsep}{4pt} 
\renewcommand{\arraystretch}{1.38} 
\scalebox{0.35}{} 
\begin{tabular}{|c|c|c|c|c|c|c|}
\hline 
 \multicolumn{6}{|c|}{$\Lambda_b\to \Lambda(\to p \pi) \mu^{+} \mu^{-}$ decay ($\mu$ mode)}\\
\hline 
\hline 
\multicolumn{2}{|c|}{SMEFT Couplings} & {BR$\times 10^{-7}$} & {$F_L$} &{$A_{FB}^{\mu}$} &{$R_{\Lambda}$} \\
  \hline 
  \hline

 \hline 
    \multirow{6}{*}{ $(\widetilde{c}_{ql}^{(3)},\widetilde{c}_{Z}^{'})$}  & $1.1-6.0$&0.628 &  0.705 &  0.070&0.807 \\
                                                    &  & [0.076, 2.650] & [0.104, 0.896] & [-0.274, -0.274]& [0.040, 2.958]   \\
                                                  
  \cline{2-6}
                                                    &$15.0-q^2_{max}$ &1.934 &  0.369&   0.050   &0.523    \\
                                                    & & [0.157, 10.021] &  [0.278, 0.511] & [-0.403, 0.064]  & [0.012, 4.074]   \\

  \hline 
     \multirow{6}{*}{ $(\widetilde{c}_{Z},\widetilde{c}_{Z}^{\prime})$} & $1.1-6.0$& 0.636 &  0.829 &  0.030 &  0.817  \\
                                                    &  & [0.178, 1.989] & [0.656, 0.923]  &  [-0.099, 0.122] & [0.116, 2.405] \\
                                                   
  \cline{2-6}
                                                    &$14.2-q^2_{max}$ &2.739 &  0.350  & 0.012  &0.741    \\
                                                    & & [1.190, 8.196] & [0.282, 0.480] & [-0.410, 0.237] & [0.105, 3.197] \\ 
                                  
    \hline
     \multirow{6}{*}{  $(\widetilde{c}_{ql}^{(1)}+\widetilde{c}_{ql}^{(3)},\widetilde{c}_{Z}^{\prime})$} & $1.1-6.0$&  0.652 &  0.708 &  0.068 &0.838  \\
                                                    &  & [0.076, 2.656] & [0.101, 0.896] & [-0.275, 0.246]  & [0.040, 2.966]  \\
                                                   
  \cline{2-6}
                                                    &$15.0-q^2_{max}$ &  2.027 &  0.368  & 0.050  &0.548    \\
                                                    & & [0.158, 10.026] & [0.278, 0.510] &  [-0.403, 0.064] & [0.012, 4.120]    \\

\hline 
 \end{tabular}
 \caption{The branching ratio~(BR), longitudinal polarization fraction $F_L$, lepton forward backward asymmetry $A_{FB}^{\mu}$ and the
ratio of branching ratio $R_{\Lambda}$ for the $\Lambda_b\to \Lambda(\to p \pi) \mu^{+} \mu^{-}$ decay mode in case of few selected
$2D$ NP scenarios.}
\label{Lmumu_np_1}
 \end{table}

\begin{table}[htbp]
\centering
\setlength{\tabcolsep}{8pt} 
\renewcommand{\arraystretch}{1.3} 
\scalebox{0.45}{}
\resizebox{\columnwidth}{!}{
\begin{tabular}{|c|c|c|c|c|c|c|c|c|c|}
\hline 
 \multicolumn{10}{|c|}{$\Lambda_b\to \Lambda(\to p \pi) \mu\mu $ decay}\\
\hline 
\hline 
SMEFT Couplings & {$q^2$ bin} &{$K_{1c}$} & {$K_{1cc}$} &{$K_{1ss}$}   &{$K_{2c}$} & {$K_{2cc}$} &{$K_{2ss}$}
 &{$K_{4sc}$}&{$K_{4s}$}\\
  \hline 
  \hline 

      \multirow{4}{*}{ $(\widetilde{c}_{ql}^{(3)},\widetilde{c}_{Z}^{'})$}  & $1.1-6.0$& 0.046  & 0.148  & 0.426 & -0.009&   0.018 &  0.004  
      &  0.018 & -0.084\\
                                                      &  & [-0.183, 0.164] & [0.052, 0.448] 
                                                      & [0.276, 0.474] & [-0.107, 0.119]
                                                      & [-0.280, 0.062] & [-0.158, 0.027]
                                                      & [-0.043, 0.067] & [-0.213, 0.207]\\

    \cline{2-10}
                                                      &$15.0-q^2_{max}$ &0.033  & 0.315 &  0.342 &  0.023 &  0.144  & 0.072   &   0.043 &  0.000\\
                                                      && [-0.268, 0.043] & [0.245, 0.361] 
                                                      & [0.319, 0.378] & [-0.073, 0.212] 
                                                      & [-0.073, 0.212] & [-0.164, 0.156] 
                                                      & [-0.087, 0.071] & [-0.061, 0.051]
                                                     \\

     \hline

      \hline 
       \multirow{4}{*}{ $(\widetilde{c}_{Z},\widetilde{c}_{Z}^{\prime})$} & $1.1-6.0$& 0.020 &  0.085 &  0.457 &  0.012 &  0.023  & 0.009& 
       0.020  & 0.098  \\
                                                       && [-0.066, 0.082] & [0.172, 0.038] 
                                                       & [0.414, 0.481] & [-0.028, 0.049] 
                                                       & [-0.080, 0.066] & [-0.040, 0.030] 
                                                       &
                                                       [-0.062, 0.099] & [-0.151, 0.145] 
                                                       \\

     \cline{2-10}
                                                      &$15.0-q^2_{max}$ &0.008  & 0.325  & 0.338 & -0.132&   0.013  & 0.006  &  -0.000 &  0.197\\                                                 
                                                      && [-0.273, 0.158] & [0.260, 0.359] &
                                                      [0.321, 0.370]& [-0.189, 0.212]  
                                                      & [0.212, 0.043] & [-0.081, 0.021] 
                                                    & [-0.060, 0.010] & [-0.200, 0.200]
                                                   \\ 
                                          
  \hline

      \hline
       \multirow{4}{*}{  $(\widetilde{c}_{ql}^{(1)}+\widetilde{c}_{ql}^{(3)},\widetilde{c}_{Z}^{\prime})$} & $1.1-6.0$ &0.046  & 0.146 &  0.427
       & -0.008 &  0.019  & 0.005   & 0.018 & -0.083 \\
                                                      && [-0.183, 0.164] & [0.052, 0.052] & [0.275, 0.474]
                                                      & [-0.108, 0.119] & [-0.280, 0.063] & [-0.159, 0.028] 
                                                      & [-0.043, 0.067] & [-0.213, 0.207] \\
                                                   
    \cline{2-10}
                                                      &$15.0-q^2_{max}$ & 0.033 &  0.316 &  0.342  & 0.023  & 0.145&   0.072   
                                                      &  0.043&  -0.000 \\
                                                      &  &[-0.268, 0.043] & [0.245, 0.361] & [0.319, 0.378] 
                                                      & [-0.072, 0.212] & [-0.164, 0.156] & [-0.082, 0.078]
                                                      & [-0.061, 0.051] &
                                                      [-0.177, 0.179]  \\

 \hline 
 \end{tabular}}
 \caption{Angular observables $K_i$ for the $\Lambda_b\to \Lambda(\to p \pi) \mu^{+} \mu^{-}$ decay mode
in case of few selected $2D$ NP scenarios.}
\label{Lmumu_np_2}
 \end{table}
\begin{itemize}
 \item \textbf{BR} :
 In case of $\Lambda_b\to \Lambda^*(\to p{K}^-)\mu^+\mu^-$ decay, branching ratio deviates from the SM prediction by $\approx 1\sigma$
 in the presence of $(\widetilde{c}_{ql}^{(1)}+\widetilde{c}_{ql}^{(3)},\widetilde{c}_{Z})$ SMEFT couplings at the low $q^2$ region, whereas,
 almost $2\sigma$ deviation is observed in the presence of $(\widetilde{c}_{ql}^{(1)}+\widetilde{c}_{ql}^{(3)},\widetilde{c}_{Z}^{'})$
coupling at the high $q^2$ region.
 In case of $\Lambda_b\to\Lambda(\to p\pi)\mu^+\mu^-$ decay channel, no significant deviation is observed at low $q^2$ region.
 However, at high $q^2$ region, there is more than $1\sigma$ deviation in presence of
 $(\widetilde{c}_{Z}, \widetilde{c}_{Z}^{\prime})$ and $(\widetilde{c}_{ql}^{(1)}+\widetilde{c}_{ql}^{(3)},\widetilde{c}_{Z}^{'})$ NP
couplings. Moreover, with $(\widetilde{c}_{ql}^{(3)},\widetilde{c}_{Z}^{'})$ NP coupling, the deviation from the SM prediction is quite
significant and it is distinguishable from the SM prediction at more than $5\sigma$.

\item \bm{$F_L$}: For the $\Lb\to \Lst(\to p{K}^-)\mu^+\mu^-$ decay channel, $F_L$ deviates from the SM prediction by $1\sigma$ in the
presence of $(\widetilde{c}_{Z},\widetilde{c}_{Z}^{\prime})$ NP couplings at the low $q^2$ region. Moreover, a significant deviation of
more than $3\sigma$ is observed in the presence of $(\widetilde{c}_{ql}^{(3)},\widetilde{c}_{Z}^{\prime})$
and $(\widetilde{c}_{ql}^{(1)}+\widetilde{c}_{ql}^{(3)}, \widetilde{c}_{Z}^{\prime})$ NP couplings.
At the high $q^2$ region, $F_L$ deviates more than $2.8\sigma$ and $1.75\sigma$ in the presence of
$(\widetilde{c}_{Z},\widetilde{c}_{Z}^{\prime})$ and $(\widetilde{c}_{ql}^{(1)}+\widetilde{c}_{ql}^{(3)}, \widetilde{c}_{Z}^{\prime})$ NP
couplings, respectively. Similarly, a deviation of more than $3.5\sigma$ is observed in the presence
of $(\widetilde{c}_{ql}^{(3)}, \widetilde{c}_{Z}^{\prime})$ NP coupling.
In case of $\Lb\to \Lst(\to p\pi)\mu^+\mu^-$ channel, a deviation of more than $1\sigma$ is observed with
$(\widetilde{c}_{ql}^{(3)},\widetilde{c}_{Z}^{\prime})$ NP coupling at both low and high $q^2$ region.

\item \bm{$A_{FB}^{\mu}$}: For the $\Lb\to \Lst(\to p{K}^-)\mu^+\mu^-$ decay, a significant deviation of more than $5\sigma$ from the
SM prediction is observed in all the three NP scenarios at both low and high $q^2$ region. For the $\Lambda_b\to\Lambda(\to p\pi)\mu^+\mu^-$
decay channel, the deviation from the SM prediction is observed to be $1\sigma$ in the presence of
$(\widetilde{c}_{ql}^{(3)},\widetilde{c}_{Z}^{\prime})$
and $(\widetilde{c}_{ql}^{(1)}+\widetilde{c}_{ql}^{(3)}, \widetilde{c}_{Z}^{\prime})$ coupling in the low $q^2$ region, whereas, at the high
$q^2$ region, a deviation of more than $10\sigma$ is observed in case of all the NP scenarios.

\item \bm{$R_{\Lambda^{(*)}}$}: We observe a deviation of more than $5\sigma$ and $10\sigma$ from the SM prediction in the ratio of branching
fractions for the $\Lb\to \Lst(\to p{K}^-)\mu^+\mu^-$ and $\Lambda_b\to\Lambda(\to p\pi)\mu^+\mu^-$ decay channels in case all the NP
scenarios at both low and high $q^2$ region.

\item \bm{$K_{1c}$} : For the $\Lambda_b\to \Lambda^* (\to p K^-)\mu^+\mu^-$ decay channel, the angular observable $K_{1c}$ deviates from the 
SM prediction at more than $5\sigma$ significance in the presence of $(\widetilde{c}_{ql}^{(3)},\widetilde{c}_{Z}^{\prime})$, 
$(\widetilde{c}_{Z}$, $\widetilde{c}_{Z}^{\prime})$, and $(\widetilde{c}_{ql}^{(1)}+\widetilde{c}_{ql}^{(3)},\widetilde{c}_{Z}^{\prime})$ 
NP couplings at the low and high $q^2$ regions. For the $\Lambda_b\to \Lambda (\to p\pi)\mu^+\mu^-$ decay channel, $1\sigma$ deviation from 
the SM prediction is observed at low $q^2$ region 
with the $(\widetilde{c}_{Z}$, $\widetilde{c}_{Z}^{\prime})$ and 
$(\widetilde{c}_{ql}^{(1)}+\widetilde{c}_{ql}^{(3)},\widetilde{c}_{Z}^{\prime})$ NP couplings, whereas,
at high $q^2$ region, a deviation of more than $10\sigma$ is observed with
$(\widetilde{c}_{Z}$, $\widetilde{c}_{Z}^{\prime})$,
$(\widetilde{c}_{ql}^{(3)},\widetilde{c}_{Z}^{\prime})$ and
$(\widetilde{c}_{ql}^{(1)}+\widetilde{c}_{ql}^{(3)},\widetilde{c}_{Z}^{\prime})$ NP couplings.

\item \bm{$K_{1cc}$} : In case of $\Lb\to \Lst(\to p{K}^-)\mu^+\mu^-$ decay channel, in the presence of 
$(\widetilde{c}_{ql}^{(3)}, \widetilde{c}_{Z}^{\prime})$ and
$(\widetilde{c}_{ql}^{(1)}+\widetilde{c}_{ql}^{(3)}, \widetilde{c}_{Z}^{\prime})$ NP couplings
the angular observable $K_{1cc}$ deviates from the SM prediction at
more than $3\sigma$ at the low $q^2$ region, whereas, it deviates more than
$10\sigma$ at the high $q^2$ region.
For the $\Lambda_b\to\Lambda(\to p\pi)\mu^+\mu^-$ decay, a deviation of more than $1\sigma$ from the SM prediction is observed at the low and 
the high $q^2$ regions with $(\widetilde{c}_{ql}^{(3)}, \widetilde{c}_{Z}^{\prime})$ and 
$(\widetilde{c}_{ql}^{(1)}+\widetilde{c}_{ql}^{(3)}, \widetilde{c}_{Z}^{\prime})$ NP couplings.
 
\item \bm{$K_{1ss}$} : In $\Lb\to \Lst(\to p{K}^-)\mu^+\mu^-$ channel, we observe a deviation of $3\sigma$
from the SM prediction in the presence of $(\widetilde{c}_{ql}^{(3)}, \widetilde{c}_{Z}^{\prime})$
and $(\widetilde{c}_{ql}^{(1)}+\widetilde{c}_{ql}^{(3)}, \widetilde{c}_{Z}^{\prime})$ NP couplings at low $q^2$ region, whereas, it deviates 
more than $5\sigma$ at high $q^2$ region.
Similarly, with $(\widetilde{c}_{ql}^{(3)},
\widetilde{c}_{Z}^{\prime})$ coupling, we observe a deviation of more than $1\sigma$ at the high $q^2$ region for the
$\Lambda_b\to\Lambda(\to p\pi)\mu^+\mu^-$ decay mode.

\item \bm{$K_{2c}$} : For the $\Lb\to \Lst(\to p{K}^-)\mu^+\mu^-$ decay, we observe a deviation of more than $5\sigma$ and $10\sigma$
in case of all the NP scenarios at low and high $q^2$ regions, respectively.
For the $\Lambda_b\to\Lambda(\to p\pi)\mu^+\mu^-$ decay channel, no significant deviation is observed at the low $q^2$ region. At the high 
$q^2$ region, however, it deviates more than $10\sigma$ in case of each NP scenarios. 

\item \bm{$K_{2cc}$} : A deviation of around $3\sigma$ and $10\sigma$ is observed in the presence of 
$(\widetilde{c}_{ql}^{(3)}, \widetilde{c}_{Z}^{\prime})$ and $(\widetilde{c}_{ql}^{(1)}+\widetilde{c}_{ql}^{(3)}, 
\widetilde{c}_{Z}^{\prime})$ couplings at the low and high $q^2$ regions, respectively for the $\Lb\to \Lst(\to p{K}^-)\mu^+\mu^-$ decay mode.
Similarly, at the low and high $q^2$ regions, a deviation of more than $2\sigma$ and $10\sigma$ is observed in case of each NP scenarios 
for the $\Lambda_b\to\Lambda(\to p\pi)\mu^+\mu^-$ decay channel.

\item \bm{$K_{2ss}$} : In $\Lb\to \Lst(\to p{K}^-)\mu^+\mu^-$ decay channel, a deviation of around $2\sigma$ from the SM prediction is 
observed in the presence of $(\widetilde{c}_{ql}^{(3)}, \widetilde{c}_{Z}^{\prime})$ coupling in the low $q^2$ region. However, in the 
high $q^2$ region, a deviation of more than $5\sigma$ is observed in the presence of the $(\widetilde{c}_{ql}^{(3)}, 
\widetilde{c}_{Z}^{\prime})$ and $(\widetilde{c}_{Z}, \widetilde{c}_{Z}^{\prime})$ couplings.
Similarly, for the $\Lambda_b\to\Lambda(\to p\pi)\mu^+\mu^-$ decay channel, a deviation of more than $2\sigma$ and $10\sigma$ is observed 
in each NP scenarios at the low and high $q^2$ regions, respectively. 

\item \bm{$K_{4ss}$} : For the $\Lb\to \Lst(\to p{K}^-)\mu^+\mu^-$ decay channel, no significant deviation is observed in the low $q^2$ 
region. However, a deviation of more than $4\sigma$ and $10\sigma$ is observed in the presence of $(\widetilde{c}_{ql}^{(3)}, 
\widetilde{c}_{Z}^{\prime})$, $(\widetilde{c}_{Z}, \widetilde{c}_{Z}^{\prime})$ and $(\widetilde{c}_{ql}^{(1)}+\widetilde{c}_{ql}^{(3)}, 
\widetilde{c}_{Z}^{\prime})$ NP couplings at the high $q^2$ region.

\item \bm{$K_{5s}$} : There is a deviation of more than $4\sigma$ in the low $q^2$ region for
the $\Lb\to \Lst(\to p{K}^-)\mu^+\mu^-$ decay channel in case of all the NP scenarios. Moreover,
at the high $q^2$ region, more than $5\sigma$ deviation is observed in the presence of $(\widetilde{c}_{ql}^{(3)}, 
\widetilde{c}_{Z}^{\prime})$ and $(\widetilde{c}_{Z}, \widetilde{c}_{Z}^{\prime})$ couplings.
 
\item \bm{$K_{4sc}$} : In the low $q^2$ region, no significant deviation is observed in $K_{4sc}$ for the 
$\Lambda_b\to\Lambda(\to p\pi)\mu^+\mu^-$ 
decay mode. However, in the high $q^2$ region, a deviation of more than $5\sigma$ is observed in the presence of 
$(\widetilde{c}_{ql}^{(3)}, \widetilde{c}_{Z}^{\prime})$, $(\widetilde{c}_{Z}, \widetilde{c}_{Z}^{\prime})$, and 
$(\widetilde{c}_{ql}^{(1)}+\widetilde{c}_{ql}^{(3)}, \widetilde{c}_{Z}^{\prime})$ NP couplings.

\item \bm{$K_{4s}$} :
For the $\Lambda_b\to\Lambda(\to pK)\mu^+\mu^-$ decay channel, a deviationof around $1\sigma$ is observed in the low $q^2$ region with
$(\widetilde{c}_{ql}^{(3)}, \widetilde{c}_{Z}^{\prime})$ and $(\widetilde{c}_{ql}^{(1)}+\widetilde{c}_{ql}^{(3)}, 
\widetilde{c}_{Z}^{\prime})$ NP couplings. However, in the high $q^2$ region, $K_{4s}$ deviates from the SM prediction by more than $5\sigma$
in each NP scenarios.

\end{itemize}
 
\begin{figure}[htbp]
\centering
\includegraphics[width=8cm,height=5cm]{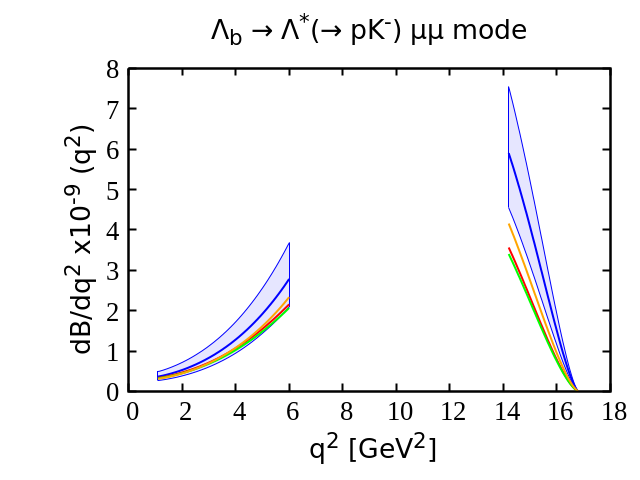}
 \includegraphics[width=8cm,height=5cm]{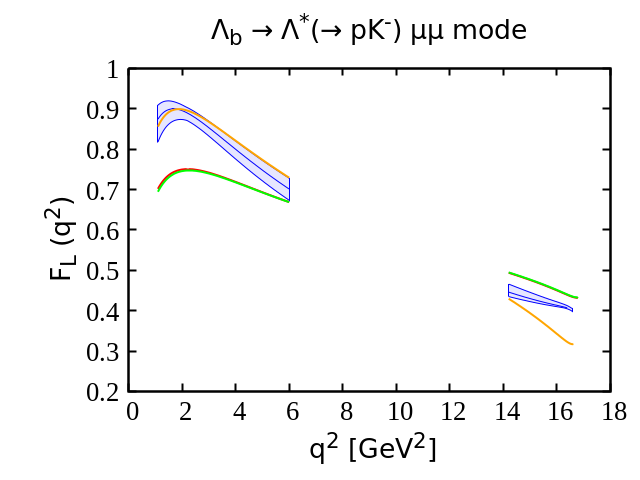}
  \includegraphics[width=8cm,height=5cm]{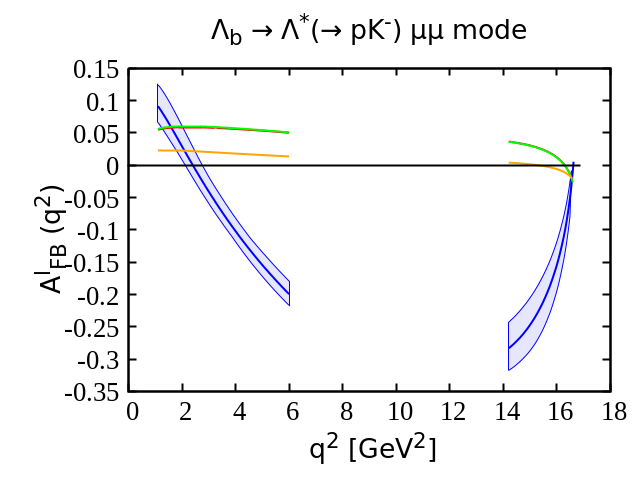}
 \includegraphics[width=8cm,height=5cm]{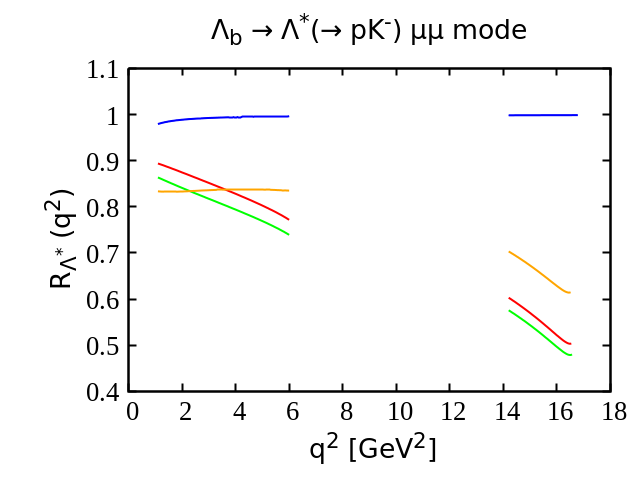}
\caption{ $q^2$ dependence of differential branching ratio~$dB/dq^2(q^2)$, longitudinal polarization fraction~$F_L(q^2)$, lepton forward
backward asymmetry~$A_{FB}^l(q^2)$ and the ratio of branching ratio~$R_{\Lambda^*}(q^2)$ for the 
$\Lambda_b\to \Lambda^* (\to p K^{-})\mu^+\mu^-$ decay mode. The SM central line and the corresponding error band is shown with blue. 
The green, orange and red lines correspond to
the best fit values of $(\widetilde{c}_{ql}^{(3)},\widetilde{c}_{Z}^{'})$,
$(\widetilde{c}_{Z},\widetilde{c}_{Z}^{\prime})$ and $(\widetilde{c}_{ql}^{(1)}+\widetilde{c}_{ql}^{(3)},\widetilde{c}_{Z}^{\prime})$, 
respectively.}

 \label{np_lst_ll}

\end{figure}

\begin{figure}[htbp]
\centering
\includegraphics[width=8cm,height=5cm]{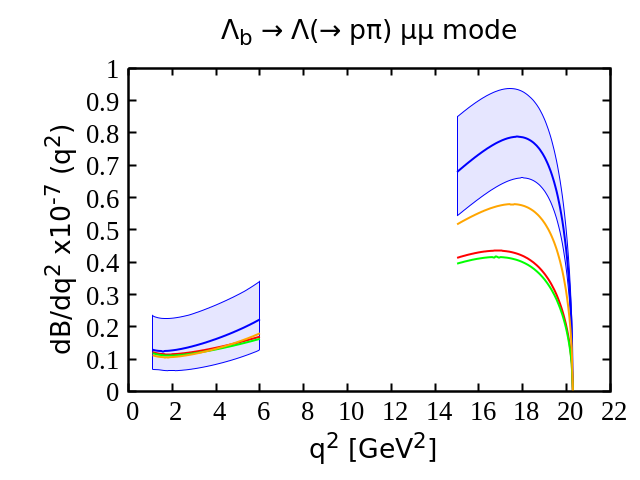}
 \includegraphics[width=8cm,height=5cm]{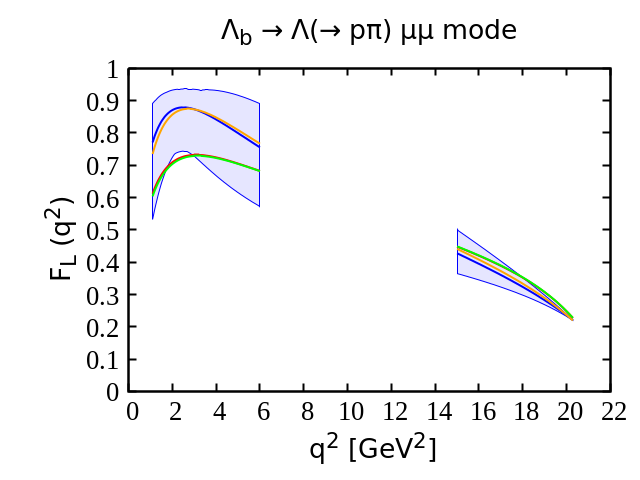}
 \includegraphics[width=8cm,height=5cm]{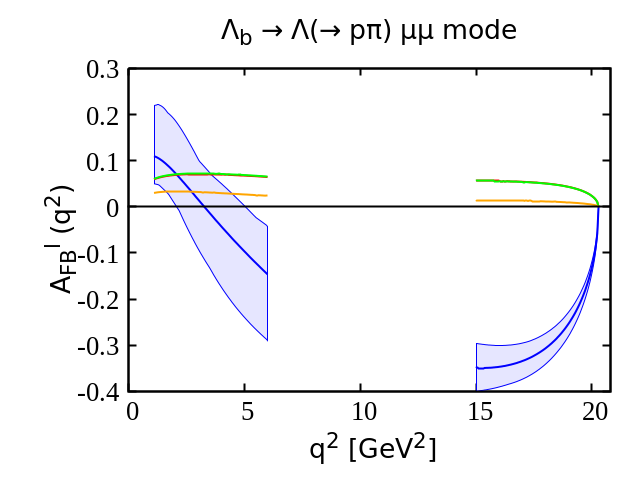}
    \includegraphics[width=8cm,height=5cm]{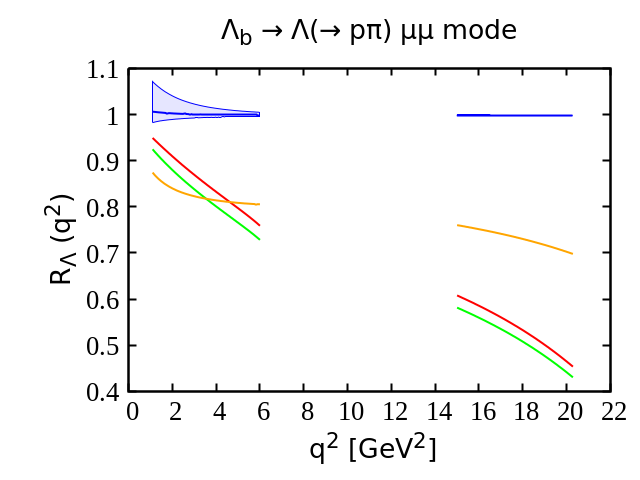}
     \caption{$q^2$ dependence of differential branching ratio~$dB/dq^2(q^2)$, longitudinal polarization fraction~$F_L(q^2)$, lepton forward 
backward asymmetry~$A_{FB}^l(q^2)$ and the ratio of branching ratio~$R_{\Lambda^*}(q^2)$ for the 
$\Lambda_b\to \Lambda (\to p \pi)\mu^+\mu^-$ decay mode. The SM central line and the corresponding error band is shown with blue. The green, 
orange and red lines correspond to
the best fit values of $(\widetilde{c}_{ql}^{(3)},\widetilde{c}_{Z}^{'})$,
$(\widetilde{c}_{Z},\widetilde{c}_{Z}^{\prime})$ and $(\widetilde{c}_{ql}^{(1)}+\widetilde{c}_{ql}^{(3)},\widetilde{c}_{Z}^{\prime})$, 
respectively.}
 \label{np_lam_ll}
 \end{figure}

In Fig.~\ref{np_lst_ll} and Fig.~\ref{np_lam_ll}, we display several $q^2$ dependent observables pertaining to 
$\Lambda_b\to\Lambda(\to pK)\mu^+\mu^-$ and $\Lambda_b\to\Lambda(\to p\pi)\mu^+\mu^-$ decay modes in the SM and in few selected NP scenarios, 
namely $(\widetilde{c}_{ql}^{(3)},\widetilde{c}_{Z}^{\prime})$,   $(\widetilde{c}_{Z},\widetilde{c}_{Z}^{\prime})$
and $(\widetilde{c}_{ql}^{(1)}+\widetilde{c}_{ql}^{(3)},\widetilde{c}_{Z}^{\prime})$, respectively. The SM central line and the
corresponding uncertainty band obtained at $95\%$ CL are shown with blue color, whereas, the effects of 
$(\widetilde{c}_{ql}^{(3)},\widetilde{c}_{Z}^{\prime})$, $(\widetilde{c}_{Z},\widetilde{c}_{Z}^{\prime})$ and 
$(\widetilde{c}_{ql}^{(1)}+\widetilde{c}_{ql}^{(3)},\widetilde{c}_{Z}^{\prime})$ NP couplings are shown with green, orange and red color 
respectively. Our main observations are as follows.

\begin{itemize}
\item {\bm{$dB/dq^2(q^2)$}}: The differential branching ratio for $\Lambda_b\to \Lambda^{*}\mu^{+}\mu^{-}$ and 
$\Lambda_b\to \Lambda \mu^{+}\mu^{-}$ decays is reduced at all $q^2$ in case of most of the NP scenarios. In 
$\Lambda_b\to \Lambda \mu^{+}\mu^{-}$ decays, the differential branching ratio is enhanced with $(\widetilde{c}_{Z},\widetilde{c}_{Z}^{'})$
NP coupling. 
All the NP scenarios are distinguishable from the SM prediction at more than $1\sigma$ in the high $q^2$ region. In the low $q^2$ region,
however, it lies within the SM error band. The deviation from the SM prediction is more pronounced in case of 
$(\widetilde{c}_{ql}^{(3)},\widetilde{c}_{Z}^{\prime})$ NP scenario.

\item \bm{$F_L(q^2)$} : For the $\Lambda_b\to \Lambda^{*}\mu^{+}\mu^{-}$ decay channel, deviation in $F_L(q^2)$ from the SM prediction is more
pronounced in case of $(\widetilde{c}_{ql}^{(3)},\widetilde{c}_{Z}^{\prime})$ and 
$(\widetilde{c}_{ql}^{(1)}+\widetilde{c}_{ql}^{(3)},\widetilde{c}_{Z}^{\prime})$ NP scenarios in the low $q^2$ region. In the high $q^2$
region, the deviation from SM prediction is more prominent in case of $(\widetilde{c}_{ql}^{(3)},\widetilde{c}_{Z}^{\prime})$ and
$(\widetilde{c}_{Z},\widetilde{c}_{Z}^{'})$ NP scenarios and they are clearly distinguishable from the SM at more than $2\sigma$ significance.
In the case of $\Lambda_b\to \Lambda\mu^{+}\mu^{-}$ decay, although a slight deviation is observed in case of 
$(\widetilde{c}_{ql}^{(3)},\widetilde{c}_{Z}^{\prime})$ NP scenario, it, however, is indistinguishable from the SM prediction.
 
\item \bm{$A_{FB}(q^2)$}: For the $\Lambda_b\to \Lambda^{*}\mu^{+}\mu^{-}$ decay channel, a significantdeviation from the SM prediction
is observed in $A_{FB}(q^2)$ in case of all the NP scenarios and they are clearly distinguishable from the SM at more than $6\sigma$ at low
and high $q^2$ regions. In the SM, we observe the zero crossing point of $A_{FB}$ at $q^2 = 2.4 \pm 0.6\,{\rm GeV^2}$ and at 
$q^2 = 16.6\pm 0.1\, {\rm GeV^2}$, respectively. With NP, there is no zero crossing of $A_{FB}$ at low $q^2$ region. However, at the high
$q^2$ region, we observe the zero crossing point at $q^2 = 15.3\, {\rm GeV^2}$ and $q^2 = 16.3\, {\rm GeV^2}$ with 
$(\widetilde{c}_{Z},\widetilde{c}_{Z}^{'})$, $(\widetilde{c}_{ql}^{(3)},\widetilde{c}_{Z}^{\prime})$ and
$(\widetilde{c}_{ql}^{(1)}+\widetilde{c}_{ql}^{(3)},\widetilde{c}_{Z}^{\prime})$ NP couplings, respectively. 
For $\Lambda_b\to \Lambda\mu^{+}\mu^{-}$ decay, in the low $q^2$ region, a slight deviation in $A_{FB}$ is observed with all the NP scenarios 
but they are indistinguishable from the SM prediction. However, at high $q^2$ region, the deviation observed is quite significant and all the
NP scenarios are distinguishable from the SM prediction at more than $10\sigma$. In the SM, a zero crossing point of $A_{FB}$ is observed at 
$q^2 = 3.3 \pm 1.5\,{\rm GeV^2}$. However, no zero crossing point is observed with NP couplings for this decay channel.

\item \bm{$R_{\Lambda^{(*)}}$ $(q^2)$}: The ratio of branching fraction $R_{\Lambda^{(*)}}(q^2)$ shows significant deviation in case of  
all the NP scenarios and it is clearly distinguishable from the SM prediction at more than $10\sigma$ significance at both low and high $q^2$ 
regions.
\end{itemize}

\begin{figure}[htbp]
\centering
 \includegraphics[width=5.9cm,height=4cm]{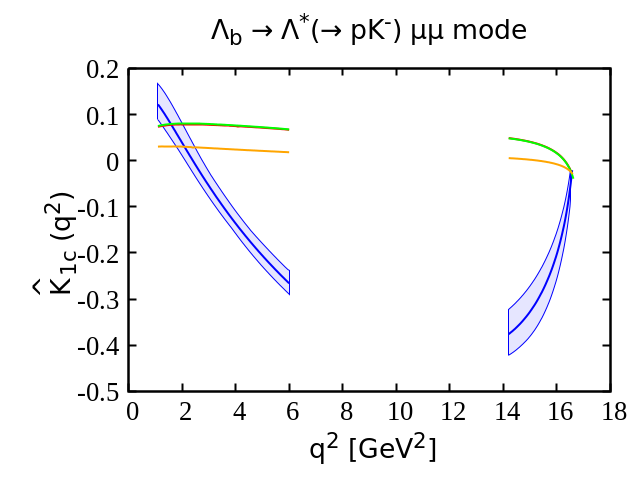}
\includegraphics[width=5.9cm,height=4cm]{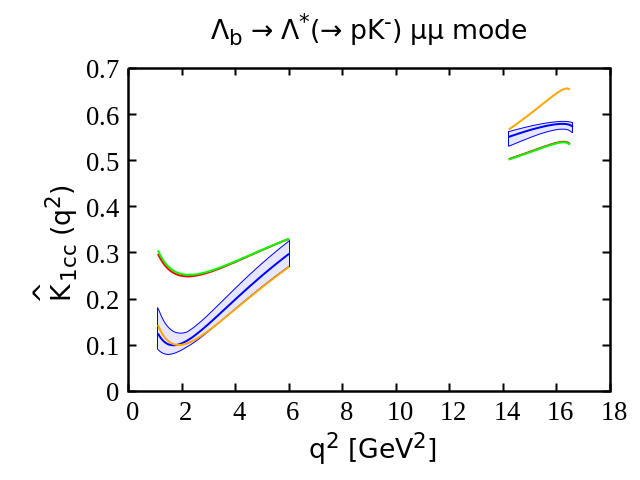}
 \includegraphics[width=5.9cm,height=4cm]{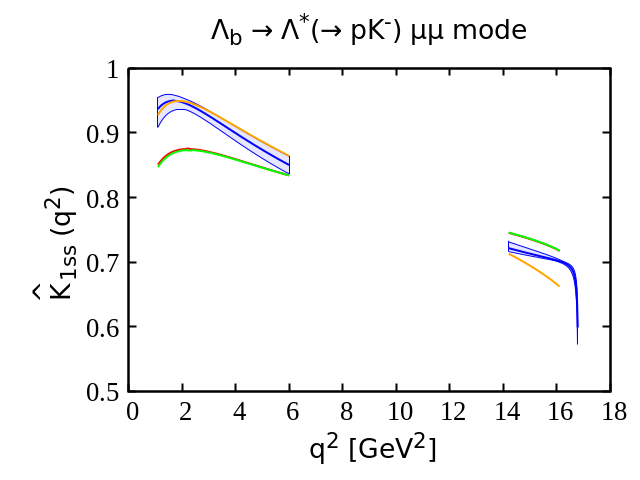}
 \includegraphics[width=5.9cm,height=4cm]{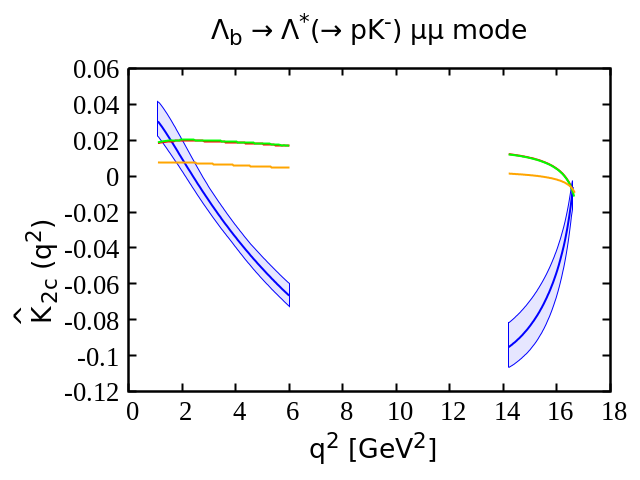}
\includegraphics[width=5.9cm,height=4cm]{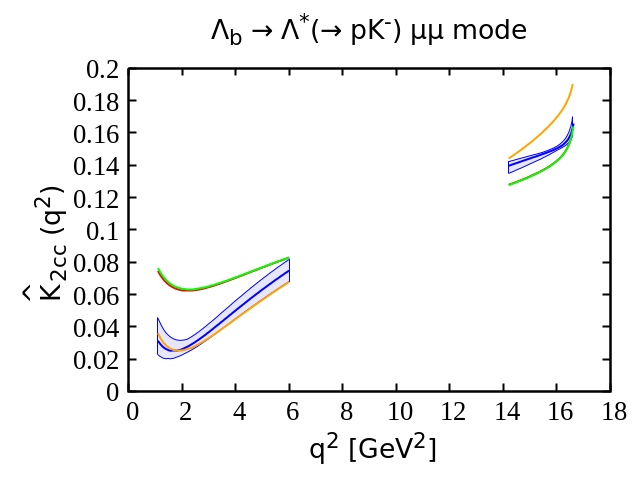}
\includegraphics[width=5.9cm,height=4cm]{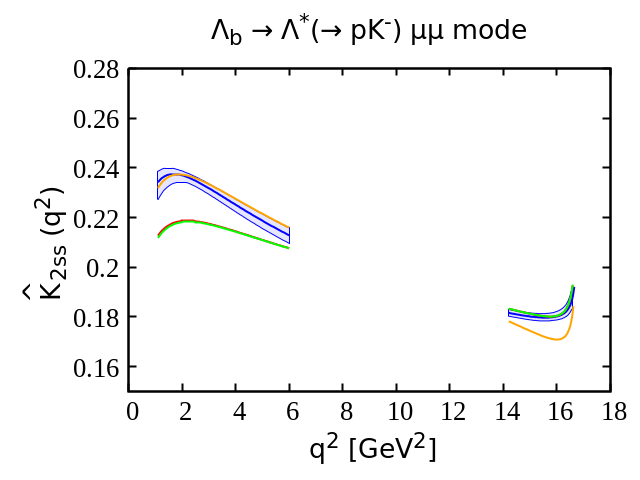}
\includegraphics[width=5.9cm,height=4cm]{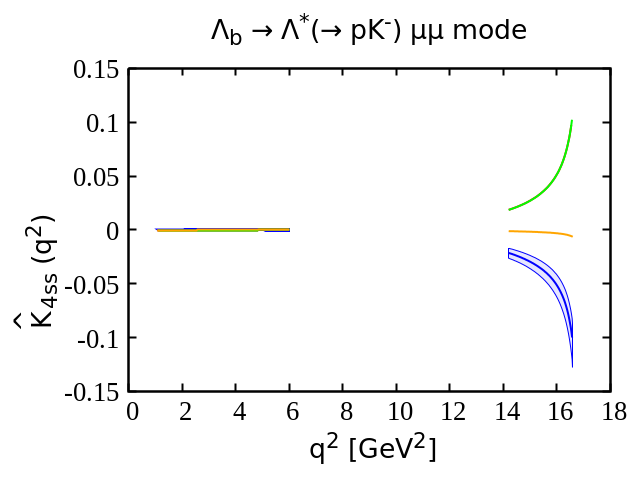}
\includegraphics[width=5.9cm,height=4cm]{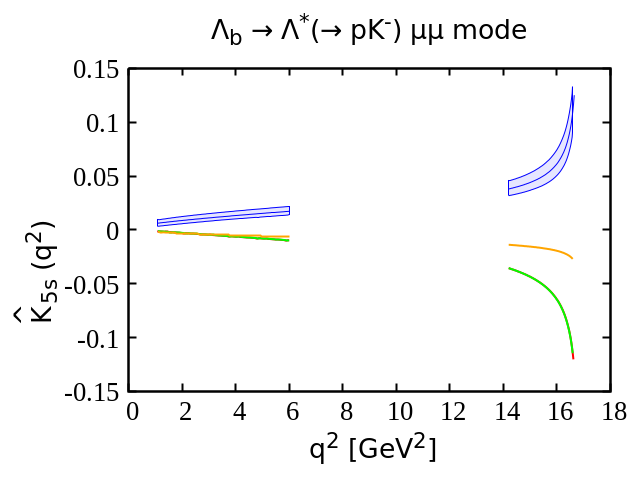}
\caption{ $q^2$ dependence of several $K$ observables for the
$\Lambda_b\to \Lambda^* (\to p K^{-})\mu^+\mu^-$ decay mode. The SM central line and the corresponding error band is shown with blue. 
The green, orange and red lines correspond to
the best fit values of $(\widetilde{c}_{ql}^{(3)},\widetilde{c}_{Z}^{'})$,
$(\widetilde{c}_{Z},\widetilde{c}_{Z}^{\prime})$ and $(\widetilde{c}_{ql}^{(1)}+\widetilde{c}_{ql}^{(3)},\widetilde{c}_{Z}^{\prime})$,
respectively.}
 \label{np_lstK_ll}
\end{figure}

\begin{figure}[htbp]
\centering
\includegraphics[width=5.9cm,height=4cm]{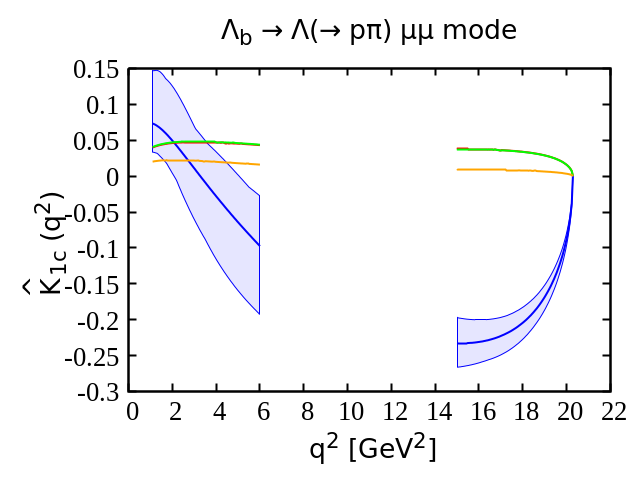}
\includegraphics[width=5.9cm,height=4cm]{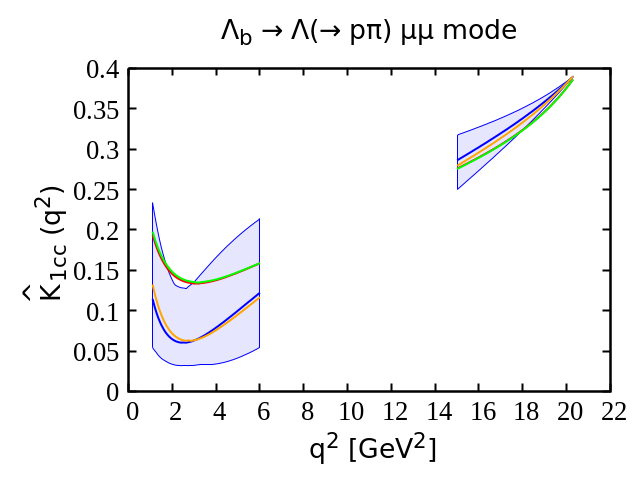}
 \includegraphics[width=5.9cm,height=4cm]{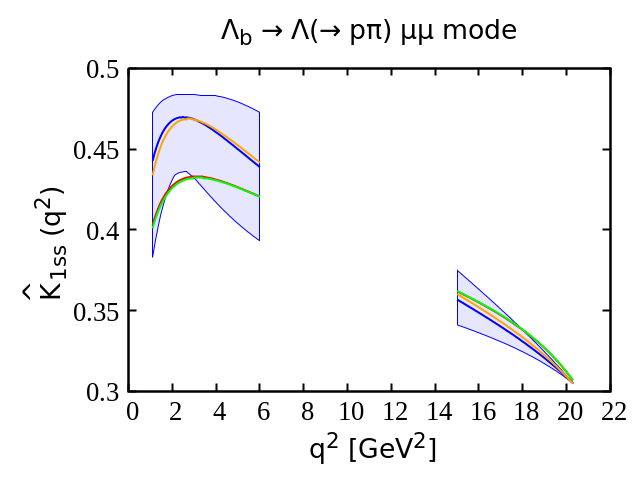}
 \includegraphics[width=5.9cm,height=4cm]{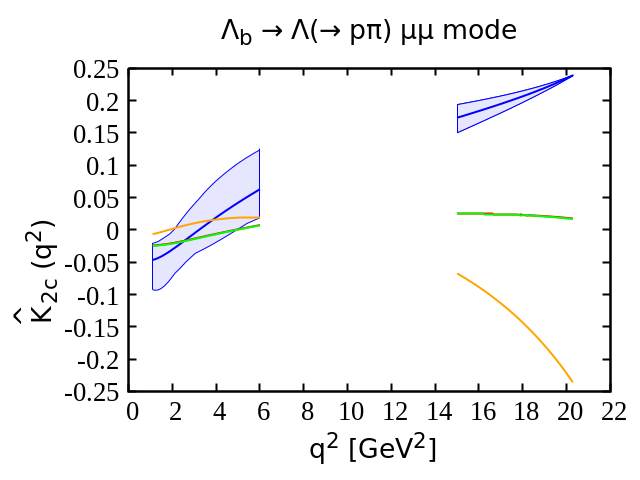}
 \includegraphics[width=5.9cm,height=4cm]{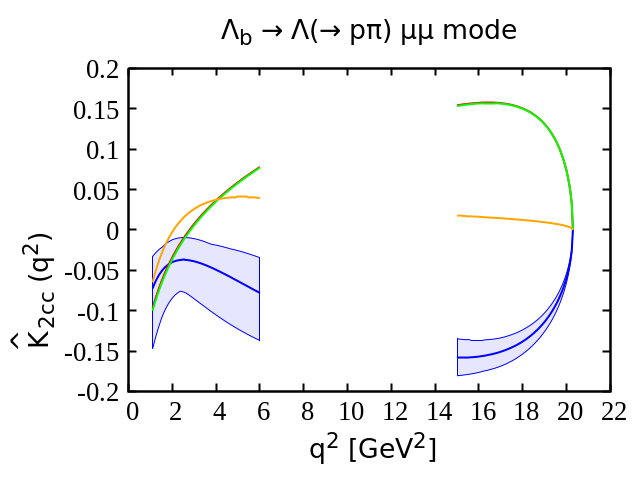}
 \includegraphics[width=5.9cm,height=4cm]{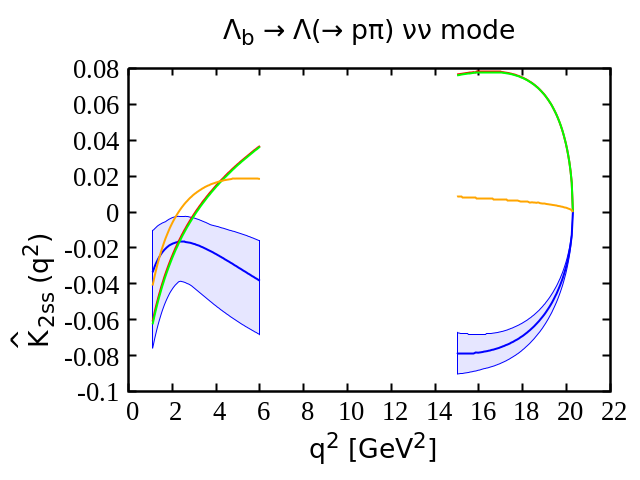}
    \includegraphics[width=5.9cm,height=4cm]{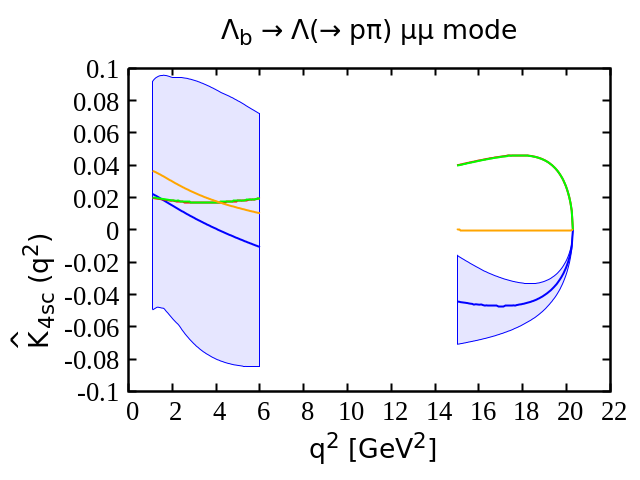}
    \includegraphics[width=5.9cm,height=4cm]{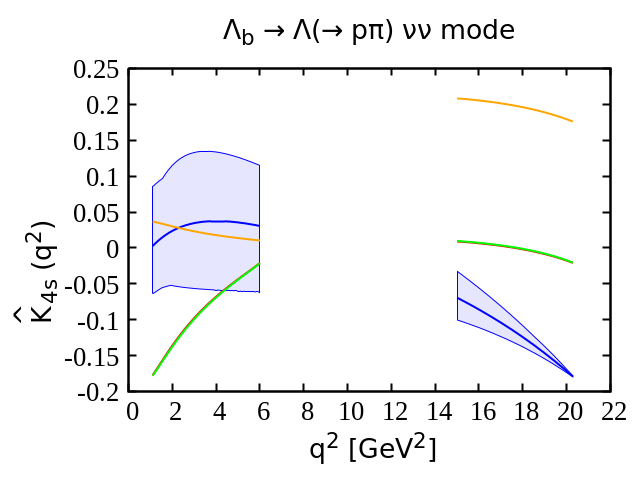}
\caption{ $q^2$ dependence of several $K$ observables for the
$\Lambda_b\to \Lambda (\to p \pi)\mu^+\mu^-$ decay mode. The SM central line and the corresponding error band is shown with blue. 
The green, orange and red lines correspond to
the best fit values of $(\widetilde{c}_{ql}^{(3)},\widetilde{c}_{Z}^{'})$,
$(\widetilde{c}_{Z},\widetilde{c}_{Z}^{\prime})$ and $(\widetilde{c}_{ql}^{(1)}+\widetilde{c}_{ql}^{(3)},\widetilde{c}_{Z}^{\prime})$,
respectively.}
 \label{np_lamK_ll}
 \end{figure}

In Fig.~\ref{np_lstK_ll} and Fig.~\ref{np_lamK_ll}, we display the NP sensitivities of several $K$ observables for the 
$\Lambda_b\to\Lambda(\to pK)\mu^+\mu^-$ and $\Lambda_b\to\Lambda(\to p\pi)\mu^+\mu^-$ decay modes in the low and high $q^2$ regions.
The SM central line and the error band is shown with blue. The green, orange and red lines correspond to NP contributions coming from
the best fit values of 
$(\widetilde{c}_{ql}^{(3)},\widetilde{c}_{Z}^{\prime})$,   $(\widetilde{c}_{Z},\widetilde{c}_{Z}^{\prime})$
and $(\widetilde{c}_{ql}^{(1)}+\widetilde{c}_{ql}^{(3)},\widetilde{c}_{Z}^{\prime})$ NP couplings of Table.~\ref{tab_fits_olddata}.

Although deviation from the SM prediction in the $K$ observables is observed in case of all the NP scenarios, it is, however, more pronounced
in case of $(\widetilde{c}_{ql}^{(3)},\widetilde{c}_{Z}^{\prime})$ and $(\widetilde{c}_{Z},\widetilde{c}_{Z}^{\prime})$ NP scenarios.  
For the $\Lambda_b\to \Lambda^{*}\mu^{+}\mu^{-}$ channel, it is observed that, irrespective of the NP contribution, the ratios
$K_{1c}/K_{2c}$, $K_{1cc}/K_{2cc}$ and $K_{1ss}/K_{2ss}$ remain independent of both short distance and long distance physics. 
For $K_{1c}$ and $K_{2c}$ the dependence on the new physics follow the same pattern as in $A_{FB}$. Similarly, for $K_{1cc}$ and $K_{2cc}$,
the dependence on the new physics follow the same pattern as in $F_L$. For the $\Lambda_b\to \Lambda\mu^{+}\mu^{-}$ channel, NP dependence of
$K_{1c}$ follows the same pattern as in $A_{FB}$. Similarly, for $K_{1cc}$ and $K_{1ss}$, the NP dependence is quite similar to that of
$F_L$. Moreover, variation of $K_{2cc}$ and $K_{2ss}$ as a function of $q^2$ looks quite similar in case of 
$\Lambda_b\to \Lambda\mu^{+}\mu^{-}$ decay channel. We observe that deviation from the SM prediction is more pronounced in case of
$(\widetilde{c}_{ql}^{(3)},\widetilde{c}_{Z}^{\prime})$ and $(\widetilde{c}_{Z},\widetilde{c}_{Z}^{\prime})$ NP scenarios.

We now proceed to discuss the effects of NP in $\Lb\to \Lst(\to p{K}^-)\nu\bar{\nu}$ and $\Lambda_b\to\Lambda(\to p\pi)\nu\bar{\nu} $
decay observables.

\subsection{Effects of SMEFT coefficients in $\Lb\to \Lst(\to p{K}^-)\nu\bar{\nu}$ and $\Lambda_b\to\Lambda(\to p\pi)\nu\bar{\nu} $ 
decay observables} 
Study of rare decays mediated via $b \to s\nu\bar{\nu}$ quark level transition can, in principle, provide complementary information
regarding NP in $b \to s\,l^+\,l^-$ transition decays. In this connection, we wish to explore the effects of NP in 
$b \to s\,l^+\,l^-$ transition decays on several observables pertaining to $\Lb\to \Lst(\to p{K}^-)\nu\bar{\nu}$ and  
$\Lambda_b\to\Lambda(\to p\pi)\nu\bar{\nu} $ decay modes. We consider three NP scenarios such as
$(\widetilde{c}_{ql}^{(3)},\widetilde{c}_{Z}^{\prime})$, $(\widetilde{c}_{Z},\widetilde{c}_{Z}^{\prime})$
and $(\widetilde{c}_{ql}^{(1)}+\widetilde{c}_{ql}^{(3)},\widetilde{c}_{Z}^{\prime})$ from Table.~\ref{tab_fits_olddata}
that best explain the anomalies present in the 
$b \to s\,l^+\,l^-$ data. Effect of these NP couplings on $\Lb\to \Lst(\to p{K}^-)\nu\bar{\nu}$ and
$\Lambda_b\to\Lambda(\to p\pi)\nu\bar{\nu} $ decay observables are listed in Table.~\ref{Lsnn}.

\begin{table}[htbp]
\centering
\setlength{\tabcolsep}{6pt} 
\renewcommand{\arraystretch}{1.2} 
\scalebox{0.45}{}
\begin{tabular}{|c|c|c||c|c|}
\hline 
& \multicolumn{2}{c}{$\Lambda_b\to \Lambda^*(\to p K^{-}) \nu\bar{\nu} $ decay} & \multicolumn{2}{|c|}{$\Lambda_b\to \Lambda(\to p \pi) \nu\bar{\nu} $ decay}\\
\hline 
SMEFT Couplings & BR$\times10^{-6}$& $F_L$ & BR$\times10^{-6}$& $F_L$ \\
\hline 

\multirow{2}{*}{ $(\widetilde{c}_{ql}^{(3)},\widetilde{c}_{Z}^{'})$} &  1.205&   0.717 & 1.007  & 0.595\\
                                                         &   [-0.001, 3.931] & [0.507, 0.731]  & [0.001, 4.243] & [0.318, 0.710]\\
\hline 
 \multirow{2}{*}{ $(\widetilde{c}_{Z},\widetilde{c}_{Z}^{\prime})$}   & 0.834  & 0.716& 0.689&   0.581\\
                                                        &     [0.006, 3.288] & [0.506, 0.730]  & [0.000, 3.624] & [0.333, 0.721]\\
\hline 
 \multirow{2}{*}{  $(\widetilde{c}_{ql}^{(1)}+\widetilde{c}_{ql}^{(3)},\widetilde{c}_{Z}^{\prime})$}    & 2.053 &  0.669  &2.060 &  0.624\\
                                                        &   [0.006, 3.288] & [0.506, 0.730]  & [1.327, 4.650] & [0.330, 0.709] \\

\hline 
\end{tabular}
 \caption{ The branching ratio~(BR) and longitudinal polarization fraction $F_L$ for the 
$\Lambda_b\to \Lambda^*(\to p K^-)\nu\bar{\nu} $ and $\Lambda_b\to \Lambda(\to p \pi) \nu\bar{\nu} $ decay modes in case of few selected
$2D$ NP scenarios.}
\label{Lsnn}
 \end{table}

Our main observations are as follows. 
 \begin{itemize}
 \item \textbf{BR} : In the $\Lb\to \Lst(\to p{K}^-)\nu\bar{\nu}$ decay channel, branching ratio deviates more than $1\sigma$ from the SM 
prediction in the presence of $(\widetilde{c}_{Z}, \widetilde{c}_{Z}^{\prime})$ and $(\widetilde{c}_{ql}^{(1)}+\widetilde{c}_{ql}^{(3)},
\widetilde{c}_{Z}^{\prime})$ NP couplings.
Similarly, in the $\Lambda_b\to\Lambda(\to p\pi)\nu\bar{\nu}$ decay channel, the branching ratio deviates more than $2\sigma$ 
in the presence of $(\widetilde{c}_{ql}^{(3)},\widetilde{c}_{Z}^{\prime})$ and $(\widetilde{c}_{Z}, \widetilde{c}_{Z}^{\prime})$ NP couplings. 
 
\item \bm{$F_L$} :  For the $\Lb\to \Lst(\to p{K}^-)\nu\bar{\nu}$ decay mode, $F_L$ shows $2\sigma$, $3.3\sigma$ and $4\sigma$ deviations 
from the SM prediction in the presence of 
$(\widetilde{c}_{ql}^{(1)}+\widetilde{c}_{ql}^{(3)},\widetilde{c}_{Z}^{\prime})$, $(\widetilde{c}_{ql}^{(3)}, \widetilde{c}_{Z}^{\prime})$ and
$(\widetilde{c}_{Z}, \widetilde{c}_{Z}^{\prime})$ NP couplings, respectively. 
Similarly, for the $\Lambda_b\to\Lambda(\to p\pi)\nu\bar{\nu} $ decay mode, $F_L$ shows deviations of around $1\sigma$ and $2\sigma$ from the 
SM prediction in presence of these NP couplings. 
  
\end{itemize} 

\begin{figure}[ht!]
\centering
\includegraphics[width=8cm,height=5cm]{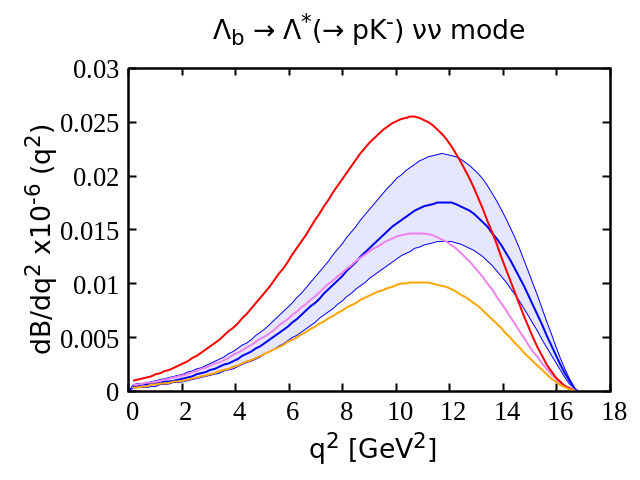}
\includegraphics[width=8cm,height=5cm]{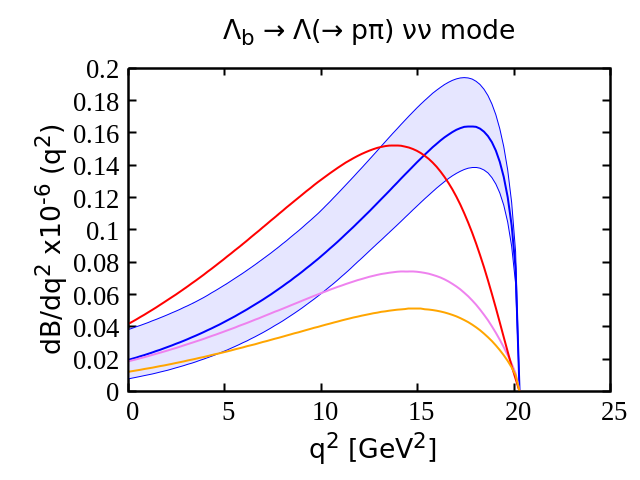}
\includegraphics[width=8cm,height=5cm]{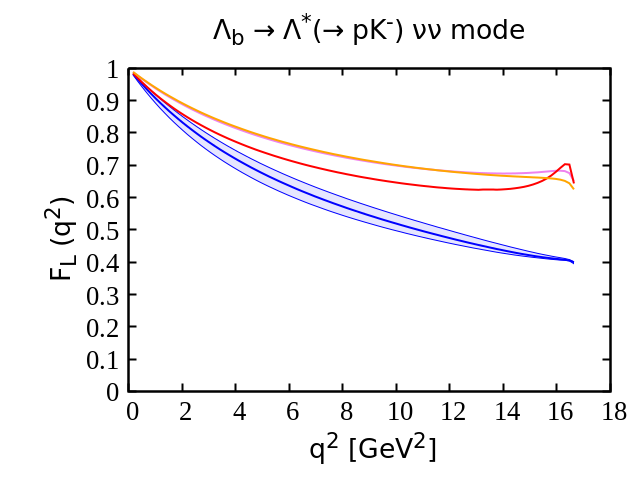}
\includegraphics[width=8cm,height=5cm]{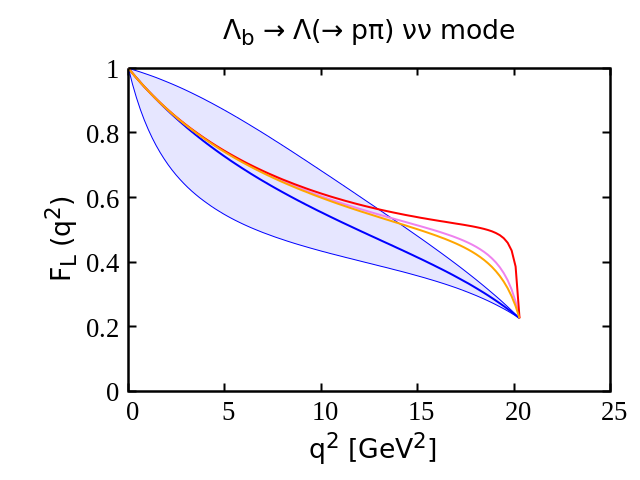}
     \caption{$q^2$ dependence of differential branching ratio~$dB/dq^2(q^2)$ and longitudinal polarization fraction~$F_L(q^2)$ for the
$\Lambda_b\to \Lambda^* (\to p K^-)\nu\bar\nu$ and $\Lambda_b\to \Lambda (\to p \pi)\nu\bar\nu$ decay mode. The SM central line and the 
corresponding error band is shown with blue. The violet, orange and red lines correspond to
the best fit values of $(\widetilde{c}_{ql}^{(3)},\widetilde{c}_{Z}^{'})$,
$(\widetilde{c}_{Z},\widetilde{c}_{Z}^{\prime})$ and $(\widetilde{c}_{ql}^{(1)}+\widetilde{c}_{ql}^{(3)},\widetilde{c}_{Z}^{\prime})$,
respectively.}
 \label{np_lst_nn}
\end{figure}

In Fig.~\ref{np_lst_nn}, we display differential branching ratio $dB/dq^2$ and longitudinal polarization fraction $F_L(q^2)$ pertaining to 
$\Lambda_b\to \Lambda^{(*)} \nu\bar{\nu}$ decay modes 
in the SM and in case of 
$(\widetilde{c}_{ql}^{(3)},\widetilde{c}_{Z}^{'})$, $(\widetilde{c}_{Z},\widetilde{c}_{Z}^{'})$ and
$(\widetilde{c}_{ql}^{(1)}+\widetilde{c}_{ql}^{(3)},\widetilde{c}_{Z}^{\prime})$ NP scenarios.
The SM central line and the corresponding uncertainty band obtained at $95\%$ CL are shown with blue color, whereas, the effects of 
$(\widetilde{c}_{ql}^{(3)},\widetilde{c}_{Z}^{'})$, $(\widetilde{c}_{Z},\widetilde{c}_{Z}^{'})$ and
$(\widetilde{c}_{ql}^{(1)}+\widetilde{c}_{ql}^{(3)},\widetilde{c}_{Z}^{\prime})$ are represented by violet, orange and red color respectively. Our observations are as follows.
 
 \begin{itemize}
\item \bm{$dB/dq^2(q^2)$} : The differential branching ratio for $\Lambda_b\to \Lambda^{*}\nu\bar{\nu}$ decays is enhanced at all $q^2$ below 
$q^2 < 12\,{\rm GeV^2}$, whereas, it is reduced at the high $q^2$ region in case of 
$(\widetilde{c}_{ql}^{(1)}+\widetilde{c}_{ql}^{(3)},\widetilde{c}_{Z}^{\prime})$ NP scenario. With 
$(\widetilde{c}_{ql}^{(3)},\widetilde{c}_{Z}^{'})$ NP coupling, the differential branching ratio lies within the SM error band except at 
$q^2 > 12\,{\rm GeV^2}$. Similarly, with $(\widetilde{c}_{Z},\widetilde{c}_{Z}^{'})$ NP coupling, it is reduced at all values of $q^2$.
The deviation from the SM prediction is more pronounced in case of 
$(\widetilde{c}_{ql}^{(1)}+\widetilde{c}_{ql}^{(3)},\widetilde{c}_{Z}^{\prime})$ and $(\widetilde{c}_{Z},\widetilde{c}_{Z}^{'})$ NP scenarios
and they are clearly distinguishable from the SM prediction at more than $2\sigma$. It should be noted that, in all the NP scenarios, the 
peak of the $q^2$ distribution appears at slightly lower value of $q^2$ than in the SM.

In case of $\Lambda_b \rightarrow \Lambda\nu\bar{\nu}$ decays, the differential branching ratio is slightly enhanced at all $q^2$ below 
$q^2 < 13\,{\rm GeV^2}$ whereas, it is reduced at the high $q^2$ region in case of  
$(\widetilde{c}_{ql}^{(1)}+\widetilde{c}_{ql}^{(3)},\widetilde{c}_{Z}^{\prime})$ NP scenario. However, it is reduced at all $q^2$
with $(\widetilde{c}_{ql}^{(3)},\widetilde{c}_{Z}^{'})$ and $(\widetilde{c}_{Z},\widetilde{c}_{Z}^{'})$ NP couplings. 
The deviation from the SM prediction is more pronounced in case of $(\widetilde{c}_{ql}^{(3)},\widetilde{c}_{Z}^{'})$ and 
$(\widetilde{c}_{Z},\widetilde{c}_{Z}^{'})$ NP scenarios and they are clearly distinguishable from the SM prediction at more than $3\sigma$.
Moreover, similar to $\Lambda_b\to \Lambda^{*}\nu\bar{\nu}$ decays, the peak of the distribution appears at slightly lower value of $q^2$ than
in the SM.

\item \bm{$F_L\,(q^2)$}: For both the decay modes, the longitudinal polarization fraction $F_L(q^2)$ is enhanced at all $q^2$ in case of all 
the NP scenarios. The deviation from the SM prediction observed in the high $q^2$ region is quite significant and they are clearly
distinguishable from the SM prediction at more than $3\sigma$. The deviation from the SM prediction is more pronounced in case of
$(\widetilde{c}_{Z},\widetilde{c}_{Z}^{'})$ NP scenario. 

\end{itemize}

\section{Conclusion}
\label{conc}
In light of anomalies observed in various $b\to s\,l^+\,l^-$ quark-level transition decays, we perform an in-depth angular analysis of 
baryonic
$\Lambda_b\to \Lambda^* (\to p K^{-})(\mu^{+}\mu^{-}, \,\nu\bar{\nu})$ and $\Lambda_b\to \Lambda (\to p \pi)(\mu^{+}\mu^{-}, \nu\bar{\nu})$ 
decays mediated
via $b\to s\,l^+\,l^-$ and $b \to s\nu\bar{\nu}$ quark level transition. 
Our main aim of this study is to explore the connections between $b\to s\,l^+\,l^-$ and $b \to s\nu\bar{\nu}$ quark level transition decays 
in a model
independent way. In this context, we use the standard model effective field theory formalism with dimension six operators that can, in 
principle, provide correlated NP effects in these decay modes. For the 
$\Lambda_b\to \Lambda^*$ form factors we use the values obtained from MCN, whereas, for the $\Lambda_b\to \Lambda$ form factors, we use the
recent results obtained from LQCD approach. We construct several NP scenarios based on NP contributions from single operators as well as from 
two different operators and try to find the scenario that best explains the anomalies present in $b\to s\,l^+\,l^-$ transition decays. To find
the best fit values of the SMEFT coefficients, we perform a naive $\chi^2$ analysis with the $b\to s\,l^+\,l^-$ data. We include total eight 
measurements in our $\chi^2$ fit. It should, however, be mentioned that, in our $\chi^2$ fit, we have not included the latest $R_K^{(*)}$ 
measurement from LHCb. It is observed that the $2D$ scenarios provide better fit to the $b\to s\,l^+\,l^-$ data than the $1D$ scenarios. More
specifically, we get much better fit with $(\widetilde{c}_{ql}^{(1)},\widetilde{c}_Z^{\prime})$,
$(\widetilde{c}_{ql}^{(3)},\widetilde{c}_{Z}^{\prime})$, $(\widetilde{c}_{Z},\widetilde{c}_{Z}^{\prime})$,
and $(\widetilde{c}_{ql}^{(1)}+\widetilde{c}_{ql}^{(3)},
\widetilde{c}_{Z}^{\prime})$ NP scenarios. The pull$_{SM}$ for these $2D$ scenarios are comparatively larger than any other scenarios.
Next we check the compatibility of our fit results with the measured values of $\mathcal B(B \to \, K^{(*)} \, \nu\,\bar{\nu})$.
It is observed that
the allowed ranges of $\mathcal{B}(B \to \, K \, \nu\,\bar{\nu})$ and $\mathcal{B}(B \to \, K^{*} \, \nu\,\bar{\nu})$ obtained with
$(\widetilde{c}_{ql}^{(3)},\widetilde{c}_{Z}^{\prime})$ and $(\widetilde{c}_{Z},\widetilde{c}_{Z}^{\prime})$ SMEFT scenarios are compatible
with the experimental upper bound. In case of $(\widetilde{c}_{ql}^{(1)},\widetilde{c}_Z^{\prime})$ and
$(\widetilde{c}_{ql}^{(1)}+\widetilde{c}_{ql}^{(3)}, \widetilde{c}_{Z}^{\prime})$ NP scenarios, although the best fit value does not
simultaneously satisfy the experimental upper bound, there still exist some NP parameter space that can, in principle, satisfy both the
constraint. 

A brief summary of our results are as follows.
\begin{itemize}
\item The differential branching ratio for the $\Lambda_b\to \Lambda^*(\to p K^{-})\mu^+\mu^-$ and 
$\Lambda_b\to \Lambda (\to p \pi)\mu^+\mu^-$ decays deviates from the SM prediction in case of all the NP scenarios and they are 
distinguishable from the SM prediction at more than $1\sigma$ in the high $q^2$ region.
Similarly, $A_{FB}(q^2)$ deviates significantly from the SM prediction in case of all the NP scenarios. For the 
$\Lambda_b\to \Lambda^*(\to p K^{-})\mu^+\mu^-$ decay mode, the zero crossing point of $A_{FB}(q^2)$ at
$q^2 = 15.3\, {\rm GeV^2}$ and $q^2 = 16.3\, {\rm GeV^2}$ with
$(\widetilde{c}_{Z},\widetilde{c}_{Z}^{'})$, $(\widetilde{c}_{ql}^{(1),(3)},\widetilde{c}_{Z}^{\prime})$ and
$(\widetilde{c}_{ql}^{(1)}+\widetilde{c}_{ql}^{(3)},\widetilde{c}_{Z}^{\prime})$ NP couplings are clearly distinguishable from the
SM zero crossing point at $q^2 = 16.6\pm 0.1\, {\rm GeV^2}$. For the $\Lambda_b\to \Lambda (\to p \pi)\mu^+\mu^-$ decays, although there
is a zero crossing point at $q^2 = 3.3 \pm 1.5\,{\rm GeV^2}$, no zero crossing point is observed with NP couplings for this decay channel.
Moreover, the ratio of branching ratio $R_{{\Lambda}^{(*)}}$ deviates significantly from the SM prediction in case of all the NP scenarios.

\item In case of $\Lambda_b\to \Lambda^* (\to p K^{-}) \,\nu\bar{\nu}$ decay, the deviation from the SM prediction in the differential 
branching ratio is more pronounced in case of
$(\widetilde{c}_{ql}^{(1)}+\widetilde{c}_{ql}^{(3)},\widetilde{c}_{Z}^{\prime})$ and $(\widetilde{c}_{Z},\widetilde{c}_{Z}^{'})$ NP scenarios
and they are clearly distinguishable from the SM prediction at more than $2\sigma$.
In case of $\Lambda_b \rightarrow \Lambda\nu\bar{\nu}$ decays, The deviation from the SM prediction is more pronounced in case of 
$(\widetilde{c}_{ql}^{(3)},\widetilde{c}_{Z}^{'})$ and
$(\widetilde{c}_{Z},\widetilde{c}_{Z}^{'})$ NP scenarios and they are clearly distinguishable from the SM prediction at more than $3\sigma$.
Similarly, $F_L$ deviates significantly from the SM prediction in the high $q^2$ region and it is clearly
distinguishable from the SM prediction at more than $3\sigma$. 
\end{itemize}

Study of $\Lambda_b\to \Lambda^* (\to p K^{-})(\mu^{+}\mu^{-}, \,\nu\bar{\nu})$
and $\Lambda_b\to \Lambda (\to p \pi)(\mu^{+}\mu^{-}, \,\nu\bar{\nu})$ mediated via $b \to s\,l\,\bar{l}$ and $b \to s\nu\bar{\nu}$ transition decays can be valuable in understanding the anomalies observed in $B$ meson decays. Our analysis can be further improved once more precise 
data on the $\Lambda_b \to \Lambda^{*}$ form factor is available from LQCD. Moreover, more precise data on 
$\mathcal{B}(B \to \, K \, \nu\,\bar{\nu})$ and $\mathcal{B}(B \to \, K^{*} \, \nu\,\bar{\nu})$ in future, can, in principle, put severe
constraint on several NP scenarios.

\FloatBarrier
\section*{Acknowledgement}
We would like to  express our gratitude to N. Rajeev for insightful and engaging discussions related to the topic addressed in this article.
We would also like to thank Diganta Das, Jaydeb Das and Stefan Meinel for their helpful exchanges regarding $\Lambda_b\to \Lambda^*$ 
transition form factors.

\bibliographystyle{ieeetr}
\bibliography{my_bib.bib}
 
\end{document}